\documentclass[journal]{IEEEtran}

\usepackage[switch]{lineno}

\usepackage{color}

\usepackage{cite}

\ifCLASSINFOpdf{}
    \usepackage[pdftex]{graphicx}
    \graphicspath{{./figures/}}
    \DeclareGraphicsExtensions{.pdf}
\else
    \usepackage[dvips]{graphicx}
    \graphicspath{{./eps/}}
    \DeclareGraphicsExtensions{.eps}
\fi
\usepackage{amssymb}
\usepackage{amsmath}
\usepackage{algorithmic}

\usepackage{array}

\usepackage{booktabs}
\usepackage{multirow}

\ifCLASSOPTIONcompsoc{}
    \usepackage[caption=false,font=normalsize,labelfont=sf,textfont=sf]{subfig}
\else
    \usepackage[caption=false,font=footnotesize]{subfig}
\fi
\usepackage{url}

\begin{document}

\bstctlcite{IEEEexample:BSTcontrol}

\title{Accurate Retinal Vessel Segmentation via\\ Octave Convolution Neural Network}

\author{Zhun~Fan,~\IEEEmembership{Senior Member,~IEEE,}
    Jiajie~Mo,
    Benzhang~Qiu,
    Wenji~Li,
    Guijie~Zhu,
    Chong~Li,
    Jianye~Hu,
    Yibiao~Rong,
    and~Xinjian~Chen\textsuperscript{*},~\IEEEmembership{Senior Member,~IEEE}
    \thanks{Z. Fan, J. Mo, B. Qiu, W. Li, G. Zhu, C. Li and J. Hu are with the Guangdong Provincial Key Laboratory of Digital Signal and Image Processing, College of Engineering, Shantou University, Shantou 515063, China. (e-mail: zfan@stu.edu.cn; jiajiemo@outlook.com; 13bzqiu@stu.edu.cn; liwj@stu.edu.cn; 16gjzhu@stu.edu.cn; 15cli@stu.edu.cn; jianyehu@outlook.com).}%
    \thanks{X. Chen and Y. Rong are with the State Key Laboratory of Radiation Medicine and Protection, Soochow University, Suzhou 215123, China. (e-mail: xjchen@suda.edu.cn; ybrong@stu.suda.edu.cn).}%
    \thanks{Asterisk indicates the corresponding author.}%
    \thanks{Source code available at: \protect\url{https://github.com/JiajieMo/OctaveUNet}}
    \thanks{This work has been submitted to the IEEE for possible publication. Copyright may be transferred without notice, after which this version may no longer be accessible.}
}
\maketitle


\begin{abstract}
    Retinal vessel segmentation is a crucial step
    in diagnosing and screening various diseases,
    including diabetes, ophthalmologic diseases, and cardiovascular diseases.
    In this paper, we propose an effective and efficient method
    for vessel segmentation in color fundus images
    using encoder-decoder based octave convolution networks.
    Compared with other convolution networks
    utilizing standard convolution for feature extraction,
    the proposed method utilizes octave convolutions
    and octave transposed convolutions for
    learning multiple-spatial-frequency features,
    thus can better capture retinal vasculatures with varying sizes and shapes.
    To provide the network the capability of learning
    how to decode multifrequency features,
    we extend octave convolution and propose a new operation
    named octave transposed convolution.
    A novel architecture of convolutional neural network,
    named as Octave UNet integrating both octave convolutions and
    octave transposed convolutions is proposed
    based on the encoder-decoder architecture of UNet,
    which can generate high resolution vessel segmentation
    in one single forward feeding without post-processing steps.
    Comprehensive experimental results demonstrate
    that the proposed Octave UNet outperforms the baseline UNet achieving
    better or comparable performance to
    the state-of-the-art methods with fast processing speed.
    Specifically, the proposed method achieves
    0.9664 / 0.9713 / 0.9759 / 0.9698 accuracy,
    0.8374 / 0.8664 / 0.8670 / 0.8076 sensitivity,
    0.9790 / 0.9798 / 0.9840 / 0.9831 specificity,
    0.8127 / 0.8191 / 0.8313 / 0.7963 F1 score,
    and 0.9835 / 0.9875 / 0.9905 / 0.9845
    Area Under Receiver Operating Characteristic (AUROC) curve,
    on DRIVE, STARE, CHASE\_DB1, and HRF datasets, respectively.
\end{abstract}



\section{Introduction}\label{sec:Introduction}
\IEEEPARstart{R}{etinal} vessel segmentation is a crucial prerequisite step
of retinal fundus image analysis because retinal vasculatures can aid
in accurate localization of many anatomical structures of retina.
Retinal vasculature is also extensively used for diagnosis assistance,
screening, and treatment planning of ocular diseases
such as glaucoma and diabetic retinopathy\cite{Srinidhi:2017eh}.
The morphological characteristics of retinal vessels
such as shape and tortuosity are important indicators for hypertension,
cardiovascular and many systemic diseases\cite{Fraz:2012bu,Vostatek:2017ei}.
These quantitative information obtained from retinal vasculature
can also be used for early detection of diabetes\cite{Heneghan:2002gr} and
progress monitoring of proliferative diabetic retinopathy\cite{Welikala:2014eq}.
Moreover, retinal vasculature can be directly visualized
by a non-inversive manner\cite{Vostatek:2017ei},
and are routinely adopted by large scale population based studies.
Furthermore, retinal vessel segmentation can be utilized
for biometric identification because the retinal vasculature
is found to be unique for each individual\cite{Marino:2006jf,Kose:2011bw}.

In clinical practice, retinal vasculature is often
manually annotated by ophthalmologists from fundus images.
This manual segmentation is a tedious, laborious, and time-consuming task
that requires skill training and expert knowledge.
Moreover, it is based on experiences and error-prone,
which lacks repeatability and reproducibility.
To reduce the workload of manual segmentation
and improve accuracy, processing speed,
and reproducibility of retinal vessel segmentation,
a tremendous amount of research efforts have been dedicated
in developing fully automated or semiautomated methods
for retinal vessel segmentation.
However, retinal vessel segmentation is a challenging task due to
various complexities of fundus images and retinal structures.
Firstly, quality of fundus images can differ
due to various imaging artifacts such as blurs, noises,
and uneven illuminations\cite{Srinidhi:2017eh,Fraz:2012bu}.
Secondly, various anatomical structures such as optic disc, macula, and fovea
are present in fundus images and complicate the segmentation task.
Additionally, the possible presence of abnormalities
such as exudates, hemorrhages and cotton wool spots
pose difficulties to retinal vessel segmentation.
Finally, one can argue that the complex nature of retinal vasculatures
presents the most significant challenge.
The shape, width, local intensity, and branching pattern
of retinal vessels vary greatly\cite{Fraz:2012bu}.
If we segment both major and thin vessels with the same technique,
it may tend to over segment one or the other\cite{Srinidhi:2017eh}.

Over the past decades, numerous retinal vessel
segmentation methods have been proposed in the
literature\cite{Srinidhi:2017eh,Fraz:2012bu,Soomro:2019ga,Vostatek:2017ei,Moccia:2018hw}.
The existing methods can be categorized into unsupervised methods and
supervised ones according to whether or not prior information such as
vessel ground truths are utilized as supervision
to guide the learning process of a vessel prediction model.
Without the need of ground truths and supervised training, most of the
unsupervised methods are rule-based, which mainly include
morphological approaches\cite{Zana:2001dw,Heneghan:2002gr,Mendonca:2006dq,Miri:2011dn,Fan:2019dn},
matched filtering methods\cite{Chaudhuri:1989ix,You:2011do,Wang:2013bk,Zhang:2016ip},
multiscale methods\cite{Ricci:2007cr,Nguyen:2013ce},
and vessel tracking methods\cite{AlDiri:2009jh,Liu:1993gu,Yin:2012gx}.

Supervised methods for retinal vessel segmentation are
based on binary pixel classification, i.e., predicting whether
a pixel belongs to vessel class or non-vessel class.
Traditional machine learning approaches involve two steps:
feature extraction and classification.
The first step involves hand crafting features
to capture the intrinsic characteristics of a target pixel.
Staal \textit{et al.}\cite{Staal:2004dd} propose a ridge based
feature extraction method that exploits the elongated structure of vessels.
Soares \textit{et al.}\cite{Soares:2006cz} utilize multiscale 2D
Gabor wavelet transformations for feature extraction.
Various classifiers such as
neural networks\cite{Marin:2011bz,Fathi:2014hj,Fan:2016ie},
support vector machines\cite{You:2011do},
and random forests\cite{Fan:2016kd,Zhang:2017kg}
are employed in conjunction with
hand crafted features extracted from local patches
for classifying the central pixel of patches.
The aforementioned supervised methods rely on application dependent
feature representations designed by domain experts,
which usually involve laborious feature design procedures
based on experiences.
Most of these features are extracted at multiple spatial scales
to better capture the varying sizes, shapes and scales of vasculatures.
However, hand crafted features may not generalize well,
especially in cases of pathological retina and complex vasculatures.

Differing from traditional machine learning approaches,
modern deep learning techniques learn
hierarchical feature representations
through multiple levels of abstraction
from fundus images and vessel ground truths automatically.
Li \textit{et al.}\cite{Li:2016jl} propose a cross modality learning framework
employing a de-noising auto-encoder for learning initial features
for the first layer of neural network.
This approach is extended by Fan and Mo in\cite{Fan:2016ie},
where a stacked de-noising auto-encoder
is used for greedy layer-wise pre-training
of a feedforward neural network.
However, these neural networks are
fully connected between adjacent layers,
which leads to problems such as over-parameterization and overfitting
for the task of vessel segmentation.
Furthermore, the quantity of trainable parameters within a
fully connected neural networks is related to the size of input images,
which leads to high computational cost
when processing high resolution images.

To address these problems, Convolutional Neural Networks (CNNs)
are employed for vessel segmentation in recent researches.
Oliveira \textit{et al.}\cite{Oliveira:2018iq} combine multiscale
stationary wavelet transformation
and Fully Convolutional Network (FCN)\cite{Shelhamer:2017bc}
to segment retinal vasculatures within local patches of fundus images.
Another popular encoder-decoder based architecture, UNet\cite{Ronneberger:2015vw},
is introduced by Antiga\cite{Antiga:2016wf}
for vessel segmentation in fundus image patches.
Alom \textit{et al.}\cite{Alom:2019id} propose the R2-UNet,
which incorporates residual blocks\cite{He:2016ib},
recurrent convolutional neural networks\cite{MingLiang:2015cm},
and the macro-architecture of UNet.
R2-UNet is designed to segment retinal vessels
in local patches of fundus images.
However, these patch based paradigms involve
cropping patches of fundus images,
processing these patches and then merging the results,
which may cause large computational overhead
without proper parallel implementation
and may have redundant computations
when the cropped patches are overlapped.
Moreover, patch based approaches do not
account for non-local correlations
when classifying the center pixels of the patches,
which may lead to failures caused by
noises or abnormalities\cite{Fu:2016fx}.

\begin{figure}[!htbp]
    \centering
    \captionsetup{width=.4\columnwidth}
    \subfloat[Fundus image]{\includegraphics[width=.49\columnwidth]{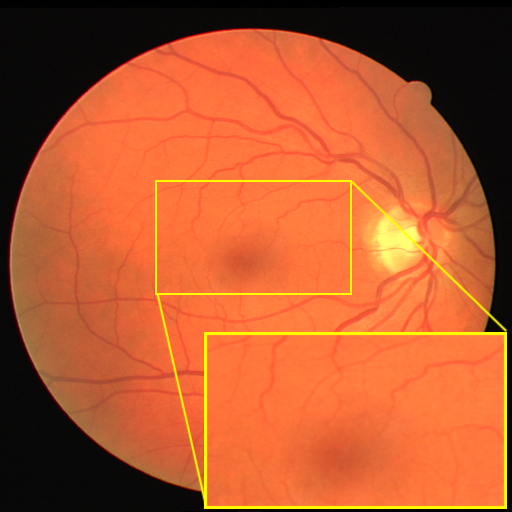}}\label{fig:thin-vessels-image}%
    \subfloat[Ground truth]{\includegraphics[width=.49\columnwidth]{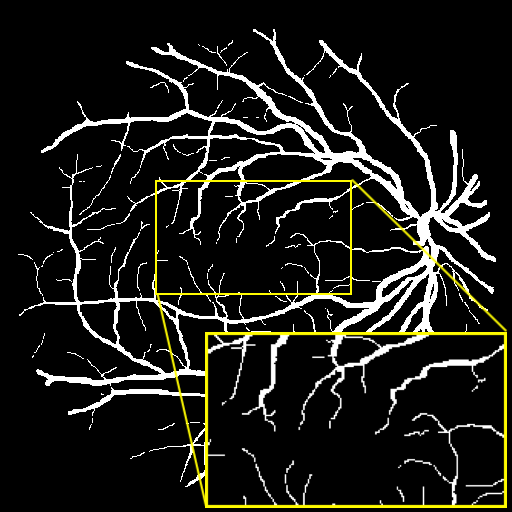}}\label{fig:thin-vessels-target}
    \caption{An example of retinal vessels with varying sizes and shapes shown in fundus image and the manually annotated ground truth.}\label{fig:thin-vessels}
\end{figure}

To overcome these issues, end-to-end approaches that process
full-sized images instead of patches are adopted.
Fu \textit{et al.}\cite{Fu:2016fx, Fu:2016fm} propose an
end-to-end approach named DeepVessel,
which is based on applying deep supervision\cite{Lee:2015tq} on
multiscale and multilevel FCN features
and adopting conditional random field formulated
as a recurrent neural network\cite{Zheng:2015cu}.
A similar deep supervision strategy is adopted
by Mo and Zhang\cite{Mo:2017io} on a deeper FCN model,
which achieves better vessel segmentation performance
than DeepVessel\cite{Fu:2016fx}.
Lei \textit{et al.}\cite{Lei:2020} propose a dense dilated network
to obtain the initial vessel segmentation, and then combine
a probability regularized walk algorithm
to address the fracture issue in the initial detection.
Although these existing methods have been successful
in segmenting major vessels,
accurately segmenting thin vessels
as shown in Fig.\ref{fig:thin-vessels}
remains a challenging problem.

In this paper, we propose an effective and efficient method
for accurately segmenting both major and thin vessels in fundus images
through learning and decoding hierarchical features
with multiple-spatial-frequencies.
The main contributions of this work are in three folds:
\begin{enumerate}
    \item Motivated by the observation that an image of vasculatures
          can be decomposed into low spatial frequency components that describe
          the smoothly changing structures such as major vessels
          and high spatial frequency components that describe
          the abruptly changing details such as thin vessels,
          we adopt octave convolution\cite{Chen:2019tp}
          for building feature encoder blocks,
          and use them to learn hierarchical multifrequency features
          at multiple levels of a neural network.
          Moreover, we demonstrate that these low- and high- frequency
          features learned have different characteristics
          and can improve the performance of the baseline model
          with only standard convolutions.
    \item For decoding these multifrequency features,
          we propose a novel operation called
          octave transposed convolution.
          This operation takes in feature maps with multiple spatial frequencies
          and restores the spatial details
          by learning multiple sets of transposed convolutional kernels.
          Decoder blocks are built upon these operations,
          and then utilized for decoding multifrequency feature maps.
    \item We also propose a novel encoder-decoder based
          neural network architecture named Octave UNet,
          which contains two main components.
          The encoder utilizes multiple aforementioned
          multifrequency encoder blocks for hierarchical
          multifrequency feature learning,
          whereas the decoder contains multiple
          aforementioned multifrequency decoder blocks
          for hierarchical feature decoding.
          Skip connections similar to those in
          the standard UNet\cite{Ronneberger:2015vw}
          are also adopted to feed additional location-information-rich
          feature maps to the decoder blocks to facilitate
          recovering spatial details and generating the
          high-resolution probability maps of vessels.
          The proposed Octave UNet can be trained in an end-to-end manner
          and deployed to produce vessel segmentation in a single forward feeding
          without the need of post-processing procedures.
          Furthermore, the Octave UNet outperforms the baseline UNet
          in terms of both segmentation performances and computational expenditure,
          while achieving better or comparable performances
          to the state-of-the-art methods on four publicly available datasets.

\end{enumerate}

The remaining sections are organized as following:
Section~\ref{sec:Method} presents the proposed method.
Section~\ref{sec:Datasets} introduces the datasets and
data augmentation technique used in this work.
In Section~\ref{sec:Experiments},
we present the training methodology and details of implementation,
along with comprehensive experimental results.
Finally, we conclude this work in Section~\ref{sec:Conclusion}.

\section{Method}\label{sec:Method}
\subsection{Multifrequency feature learning}\label{subsec:OctConv}
\begin{figure*}[!htbp]
    \centering
    \subfloat[Fundus image]{\includegraphics[width=.49\columnwidth]{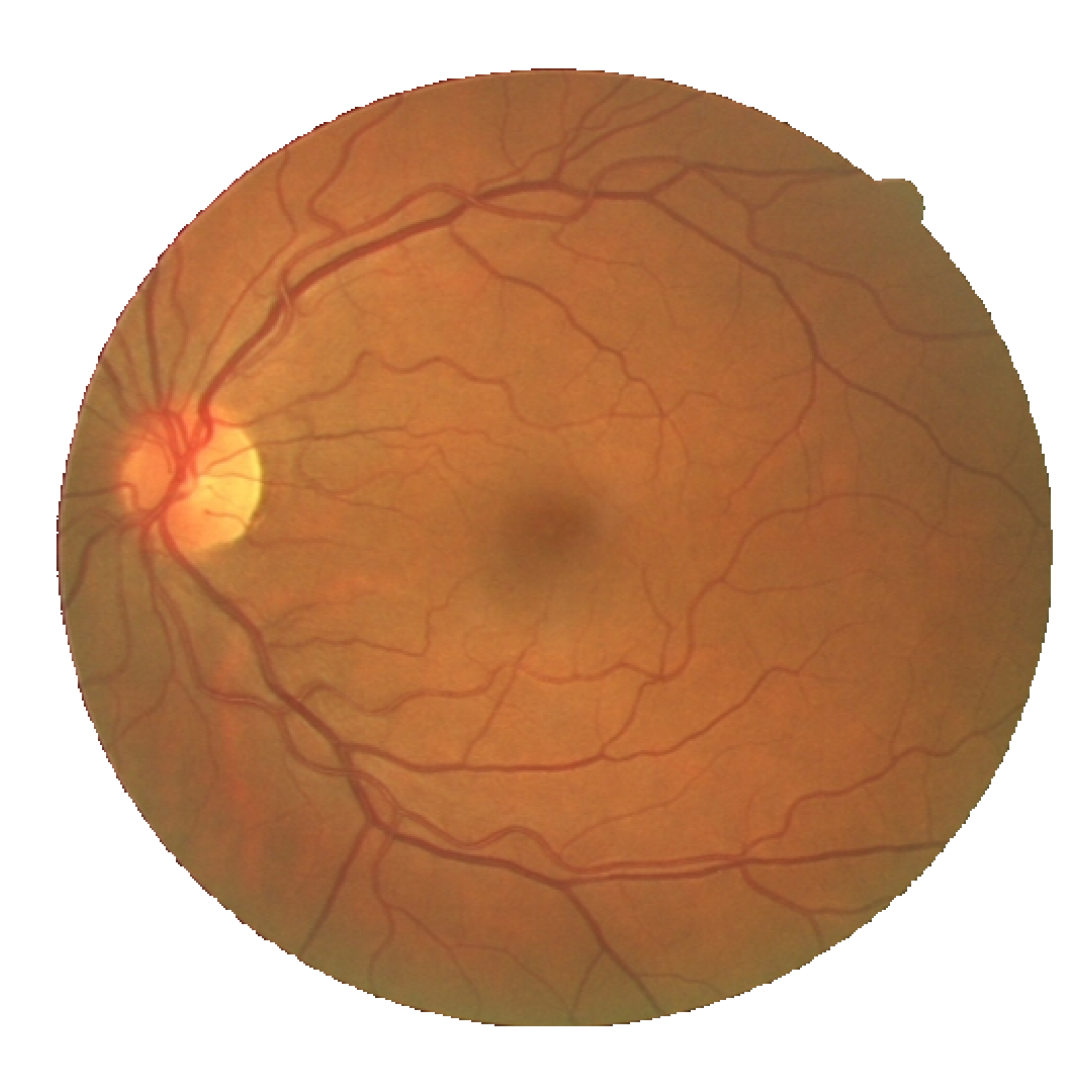}}\label{fig:org-image}%
    \subfloat[Image of vasculature]{\includegraphics[width=.49\columnwidth]{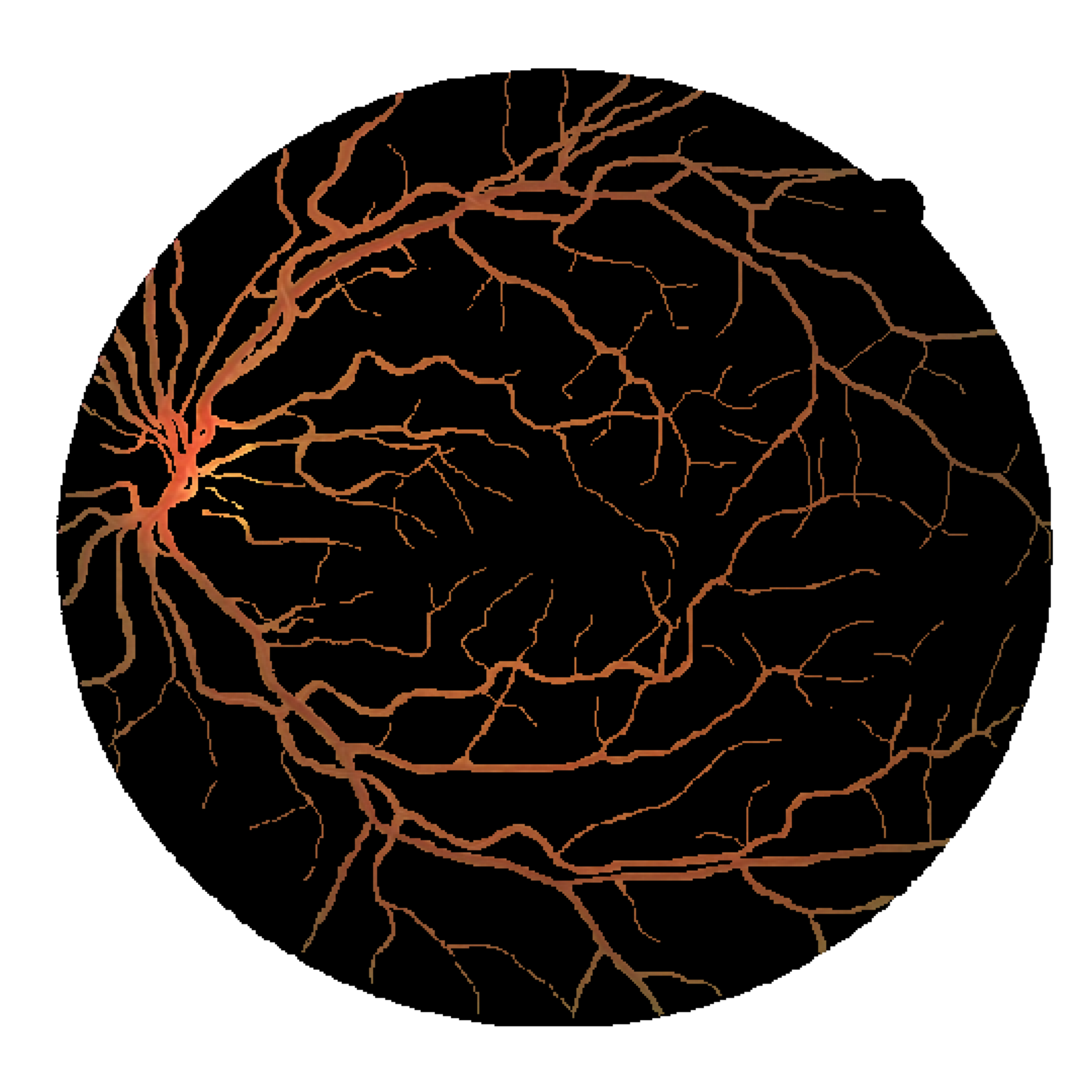}}\label{fig:vessel-image}%
    \subfloat[Low frequency component]{\includegraphics[width=.49\columnwidth]{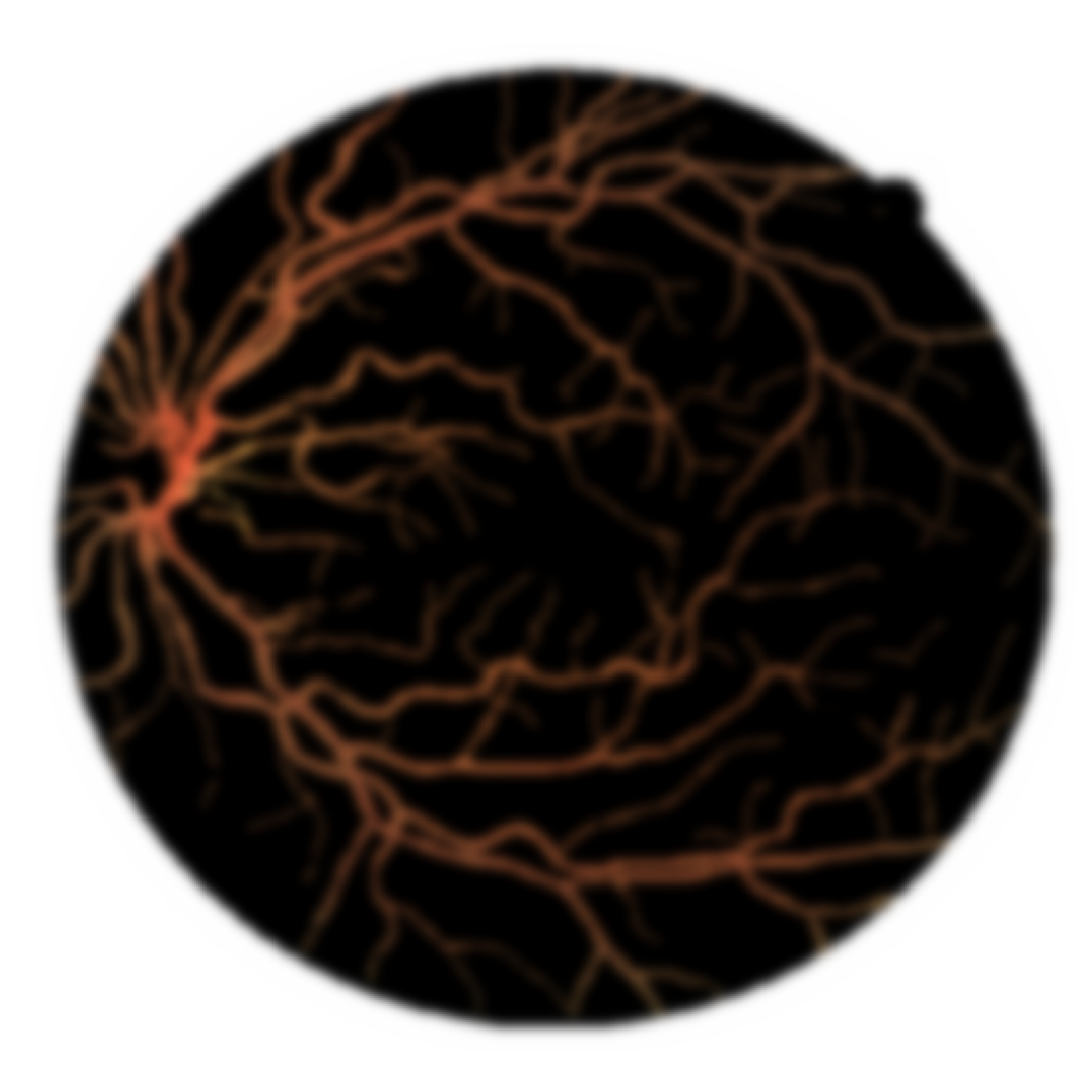}}\label{fig:vessel-low}%
    \subfloat[High frequency component]{\includegraphics[width=.49\columnwidth]{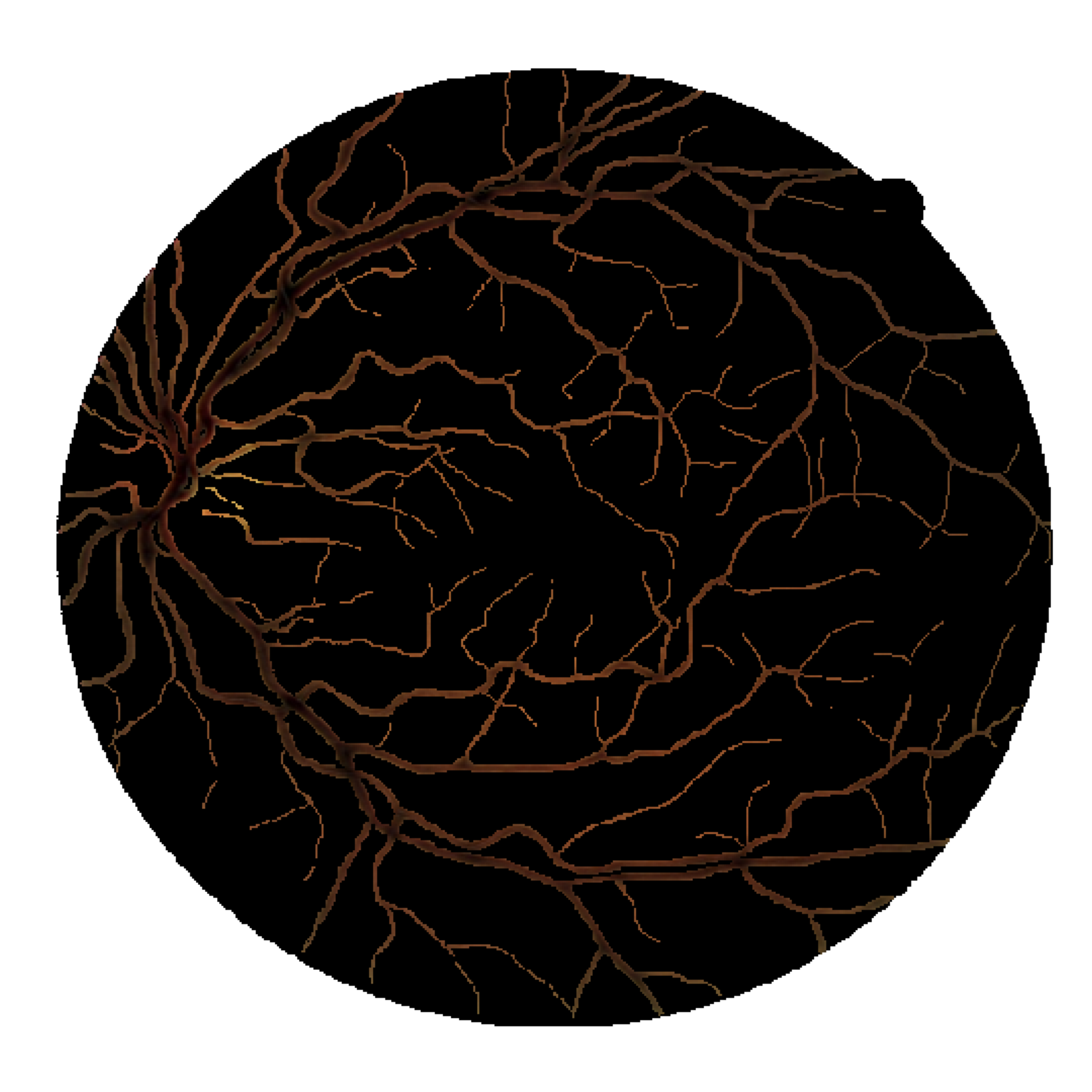}}\label{fig:vessel-high}
    \caption{The image of vasculature can be decomposed into low spatial frequency components that describe the major vascular tree and high spatial frequency components that describe the edges and minor details of thin vessels.}\label{fig:motivation}
\end{figure*}

\begin{figure}[!htbp]
    \centering
    \includegraphics[width=\columnwidth]{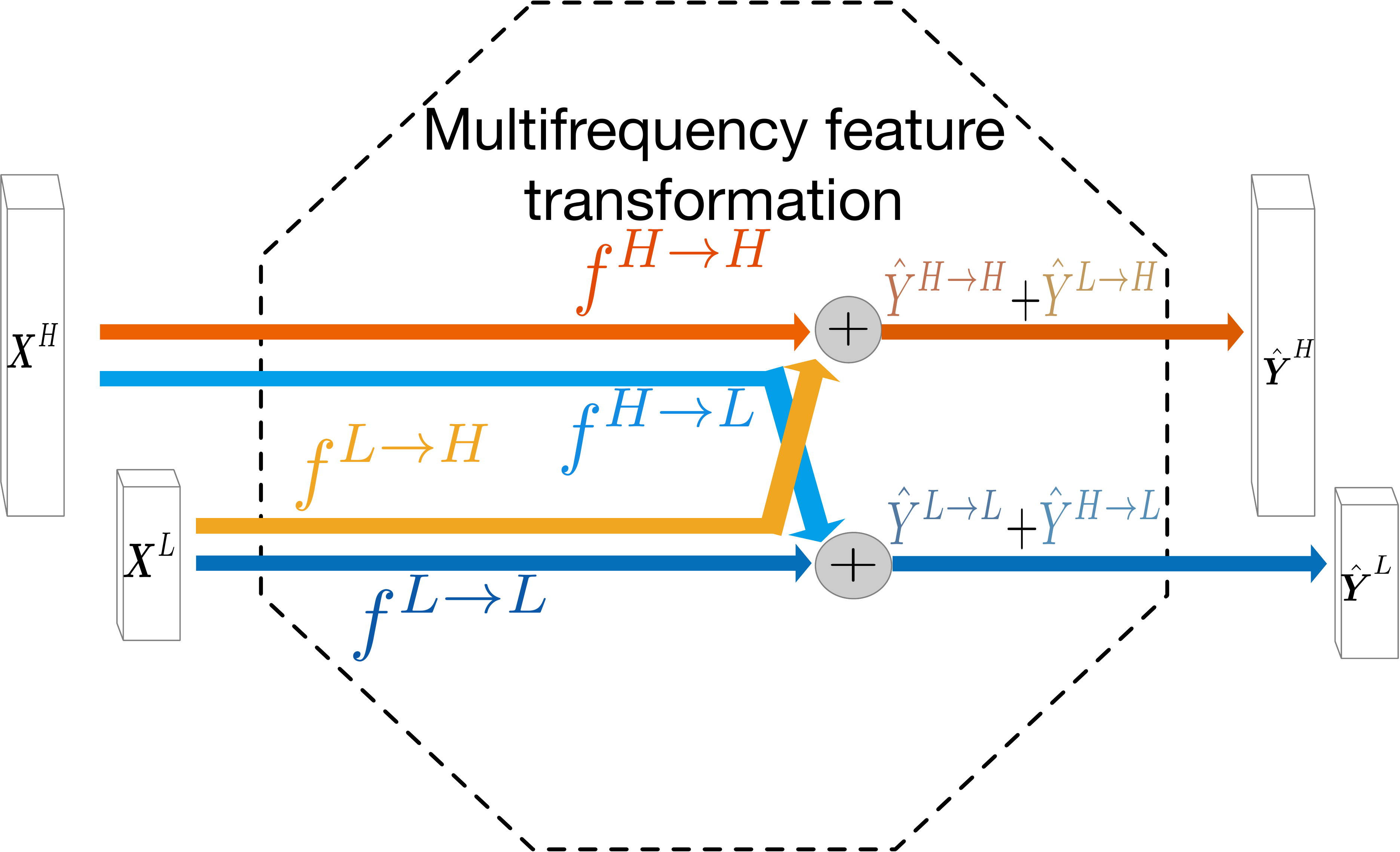}
    \caption{Computation graph of the multifrequency feature transformation of octave convolution.
        The operation mainly contains two processes of the inter-frequency information exchange (\(f^{L\rightarrow H}\) and \(f^{H\rightarrow L}\))
        and intra-frequency information update (\(f^{L\rightarrow L}\) and \(f^{H\rightarrow H}\)).
    }\label{fig:OctConv}
\end{figure}

Retinal vessel forms complex tree-like structure
with varying sizes, shapes and vessel widths.
As illustrated in Fig.\ref{fig:motivation}, the low- and high- frequency
components of retinal vasculature focus on capturing major vessels
and thin vessels, respectively.
Motivated by this observation,
we hypothesize that adopting a multi-frequency feature learning approach
may be beneficial for segmenting retinal vessels from fundus images.
Therefore, the octave convolution\cite{Chen:2019tp} is adopted
as an extractor for multifrequency features in this work.
The computational graph for multifrequency feature transformations
of the octave convolution is illustrated in Fig.\ref{fig:OctConv}.
Let \(X^{H}\) and \(X^{L}\) denote the inputs of
high- and low- frequency feature maps, respectively.
The high- and low- frequency outputs of the octave convolution are given by
\(\hat{Y}^{H}=f^{H\rightarrow H}(X^{H}) + f^{L\rightarrow H}(X^{L})\)
and \(\hat{Y}^{L}=f^{L\rightarrow L}(X^{L}) + f^{H\rightarrow L}(X^{H})\),
where \(f^{H\rightarrow H}\) and \(f^{L\rightarrow L}\)
denote two standard convolution operations
for intra-frequency information update,
whereas \(f^{H\rightarrow L}\) and \(f^{L\rightarrow H}\)
denote the process of inter-frequency information exchange.
Specifically, \(f^{H\rightarrow L}\) is equivalent to
first downsampling the input by average pooling with a scale of two
and then applying a standard convolution for feature transformation,
and \(f^{L\rightarrow H}\) is equivalent to upsampling
the output of a standard convolution by
nearest interpolation with a scale of two.

\subsection{Decoding multifrequency features}\label{subsec:OctTrConv}
On one hand, during the feature encoding process
as shown in Fig.\ref{fig:feature},
while the spatial dimensions of the feature maps reduce gradually,
the feature maps lose spatial details step by step.
This compression effect forces
the kernels to learn more discriminative features
with higher levels of abstraction.
On the other hand, given only the
multifrequency feature extraction is insufficient to
perform dense pixel classification for retinal vessel segmentation.
A process of decoding feature maps to recover spatial details
and generating high resolution probability maps of vessels is needed.
One naive way of achieving this is to use bilinear interpolation,
which unfortunately lacks the capability of
learning the decoding transformation
as possessed by the transposed convolution.
However, simply using transposed convolutions for multiple layers
ignores information exchanges between frequencies within each layer,
which may limit the capability of the model to
capture more topological relationships and
reconstruct segmentation results.

To address these issues, we extend octave convolution
and propose a novel operation named octave transposed convolution,
which provides the capability of learning suitable mappings
for decoding multifrequency features.
This operation takes in feature maps with multiple spatial frequencies
and restores their spatial details
by learning a set of transposed convolution kernels for
intra-frequency information update and
inter-frequency information exchange.

The multifrequency feature transformation of octave transposed convolution
is similar to the transformation of octave convolution
as shown in Fig.\ref{fig:OctConv}.
Let \(X^{H}\) and \(X^{L}\) denote the inputs of
high- and low- frequency feature maps, respectively.
The high- and low- frequency outputs of
the octave transposed convolution are given
by \(\hat{Y}^{H}=f^{H\rightarrow H}(X^{H}) + f^{L\rightarrow H}(X^{L})\)
and \(\hat{Y}^{L}=f^{L\rightarrow L}(X^{L}) + f^{H\rightarrow L}(X^{H})\),
where \(f^{H\rightarrow H}\) and \(f^{L\rightarrow L}\)
denote the intra-frequency information update,
whereas \(f^{H\rightarrow L}\) and \(f^{L\rightarrow H}\)
denote the inter-frequency information exchange.

Specifically, let \(W=\{W^{H\rightarrow H}, W^{L\rightarrow L}, W^{H\rightarrow L}, W^{L\rightarrow H}\} \)
denote the octave kernels composed of a set of trainable parameters,
\(b=\{b^{H\rightarrow H}, b^{L\rightarrow L}, b^{H\rightarrow L}, b^{L\rightarrow H}\} \)
denote the biases,
\(k\) denotes the size of a square kernel,
\(\sigma(\cdot)\) denotes the non-linear activation function,
and \(\left\lfloor\cdot\right\rfloor \) denotes the floor operation.
The high- and low- frequency responses at location \((i,j)\) of the output
are given by (\ref{eq:OctTrConvH}) and (\ref{eq:OctTrConvL}), respectively.

\setlength{\arraycolsep}{0.0em}
\begin{eqnarray}
    \label{eq:OctTrConvH}
    \hat{Y}^{H}_{(i,j)} &{=}& \hat{Y}^{L\rightarrow H}_{(i,j)} + \hat{Y}^{H\rightarrow H}_{(i,j)}\nonumber \\
    &{=}& f^{L\rightarrow H}(X^{L}) + f^{H\rightarrow H}(X^{H})\nonumber \\
    &{=}&\:\sigma(\sum_{m,n}{X^{L}_{(m+\frac{k-1}{2}, n+\frac{k-1}{2})}}^T W^{L\rightarrow H}_{(\left\lfloor\frac{i}{2}\right\rfloor + m, \left\lfloor\frac{j}{2}\right\rfloor + n)} + b^{L\rightarrow H})\nonumber \\
    &&{+}\:\sigma(\sum_{m,n}{X^{H}_{(m+\frac{k-1}{2}, n+\frac{k-1}{2})}}^T W^{H\rightarrow H}_{(i+m,j+n)} + b^{H\rightarrow H})
\end{eqnarray}
\setlength{\arraycolsep}{5pt}

\setlength{\arraycolsep}{0.0em}
\begin{eqnarray}
    \label{eq:OctTrConvL}
    \hat{Y}^{L}_{(i,j)} &{=}& \hat{Y}^{H\rightarrow L}_{(i,j)} + \hat{Y}^{L\rightarrow L}_{(i,j)}\nonumber \\
    &{=}& f^{H\rightarrow L}(X^{H}) + f^{L\rightarrow L}(X^{L})\nonumber \\
    &{=}&\:\sigma(\sum_{m,n}{X^{H}_{(m+\frac{k-1}{2}, n+\frac{k-1}{2})}}^T W^{H\rightarrow L}_{(2i+m+\frac{1}{2}, 2j+n+\frac{1}{2})} + b^{H\rightarrow L})\nonumber \\
    &&{+}\:\sigma(\sum_{m,n}{X^{L}_{(m+\frac{k-1}{2}, n+\frac{k-1}{2})}}^T W^{L\rightarrow L}_{(i+m,j+n)} + b^{L\rightarrow L})
\end{eqnarray}
\setlength{\arraycolsep}{5pt}

It is worth mentioning that \(f^{L\rightarrow L}\) and \(f^{H\rightarrow H}\)
are standard transposed convolution operations,
whereas \(f^{H\rightarrow L}\) is equivalent to
first downsampling the input by a scale of two
(i.e., approximating \(X^{H}_{(2i+\frac{1}{2}, 2j+\frac{1}{2})}\)
by using average of all four adjacent locations)
and then applying standard transposed convolution.
Likewise, \(f^{L\rightarrow H}\) is equivalent to upsampling
the output of standard transposed convolution by a scale of two,
where \(X^{L}_{(\left\lfloor\frac{i}{2}\right\rfloor, \left\lfloor\frac{j}{2}\right\rfloor)}\)
is implemented with nearest neighbor interpolation.

Moreover, let \(\alpha = \frac{c_{\text{low}}}{c_{\text{low}} + c_{\text{high}}}\)
denotes the ratio of number of channels of the low-frequency feature maps,
where \(c_{\text{high}}\) and \(c_{\text{low}}\) are the number of channels of
high- and low- frequency feature maps, respectively.
When \(\alpha = 0\), the octave transposed convolution outputs
only high-frequency feature maps and
the computations related to the low-frequency features are ignored.
In this case, the octave transposed convolution
becomes a standard transposed convolution operation.
Without special mentioning, all hyper-parameters \(\alpha \)
are set to 0.5 in this work.

\subsection{Octave UNet}\label{subsec:OctaveUNet}
\begin{figure}[!htbp]
    \centering
    \includegraphics[width=0.785\columnwidth]{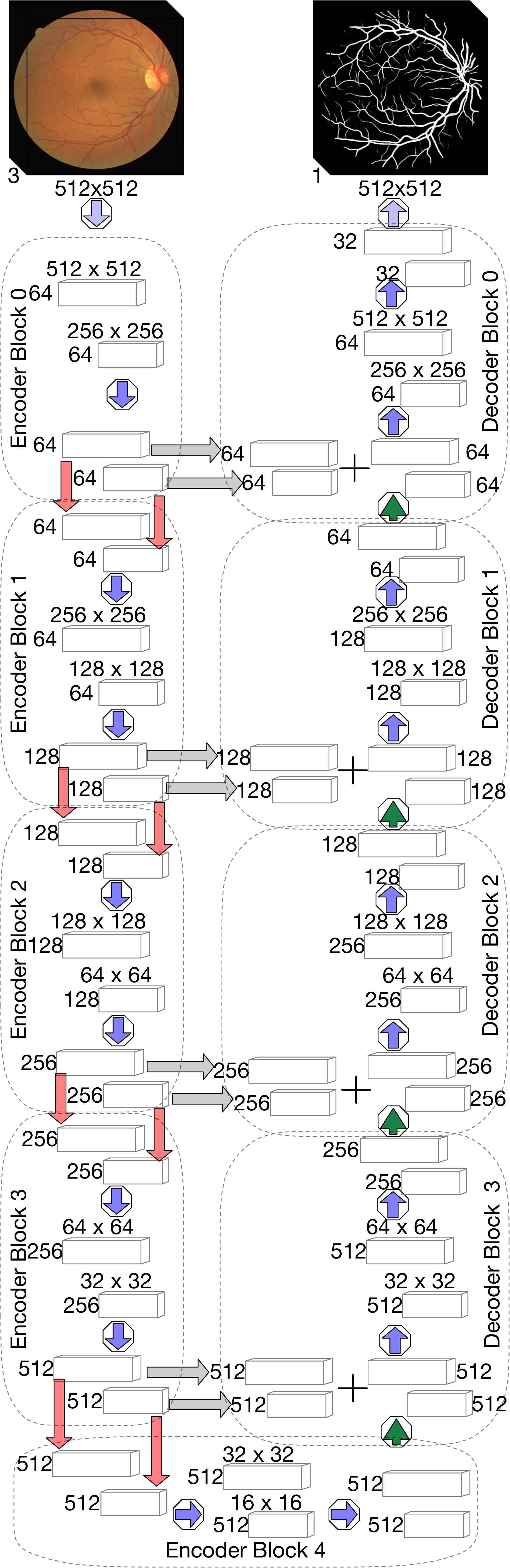}
    \caption{Detailed network architecture of the Octave UNet.
        Feature maps are denoted as cubics with number of channels on the side and spatial dimensions on the top.
        The spatial dimensions of feature maps remain the same within an encoder or decoder block.
        An octave convolutions and an octave transposed convolutions are denoted by a blue and green arrow within an octagon, respectively.
        The red arrows denote max pooling operations that downsample inputs by a scale of two.
        The gray arrows and plus signs denote skip connections that copy and concatenate feature maps.
    }\label{fig:OctaveUNet}
\end{figure}

\begin{figure*}[!htbp]
    \centering
    \subfloat[Encoder 0 (high)]{\includegraphics[width=.33\columnwidth]{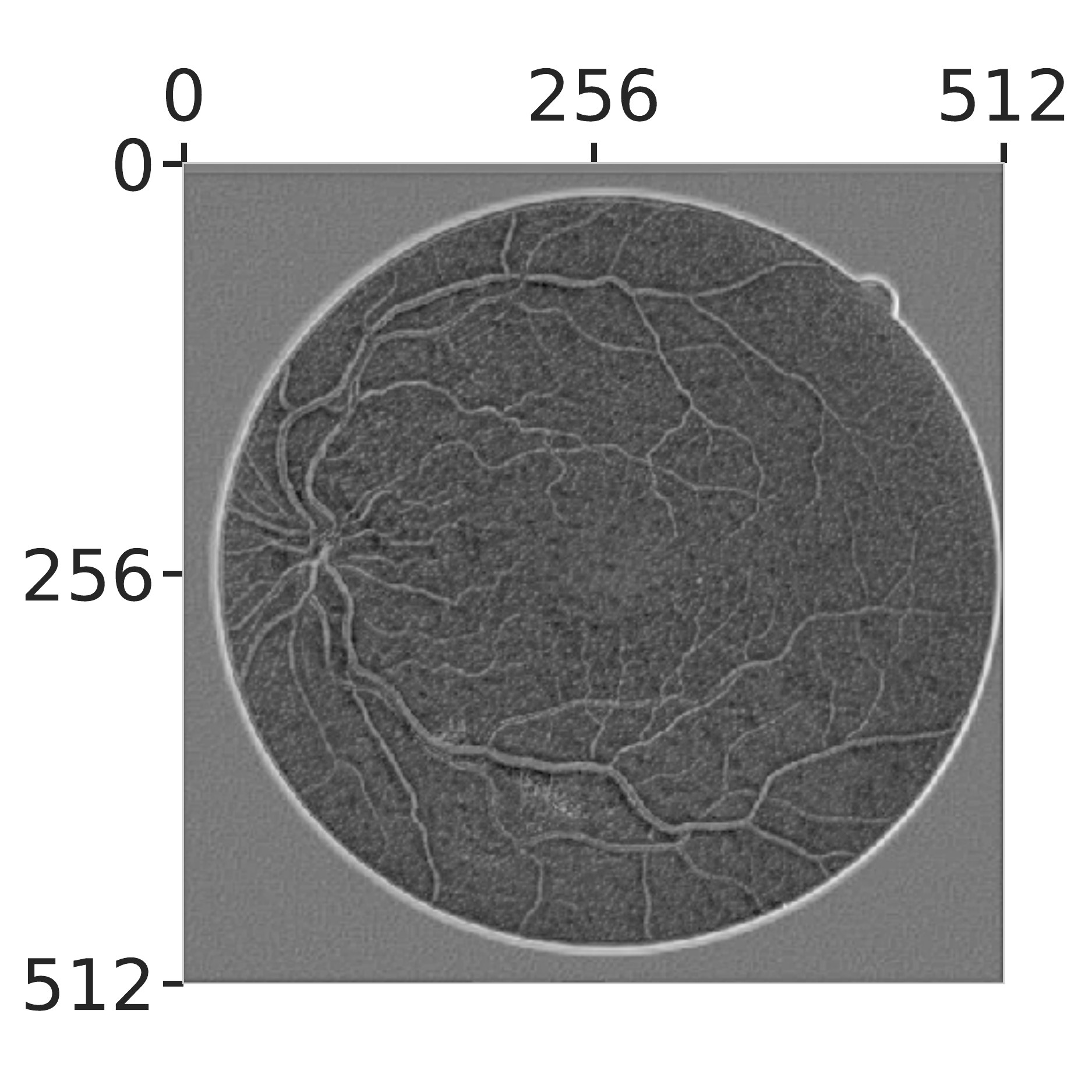}}%
    \subfloat[Encoder 1 (high)]{\includegraphics[width=.33\columnwidth]{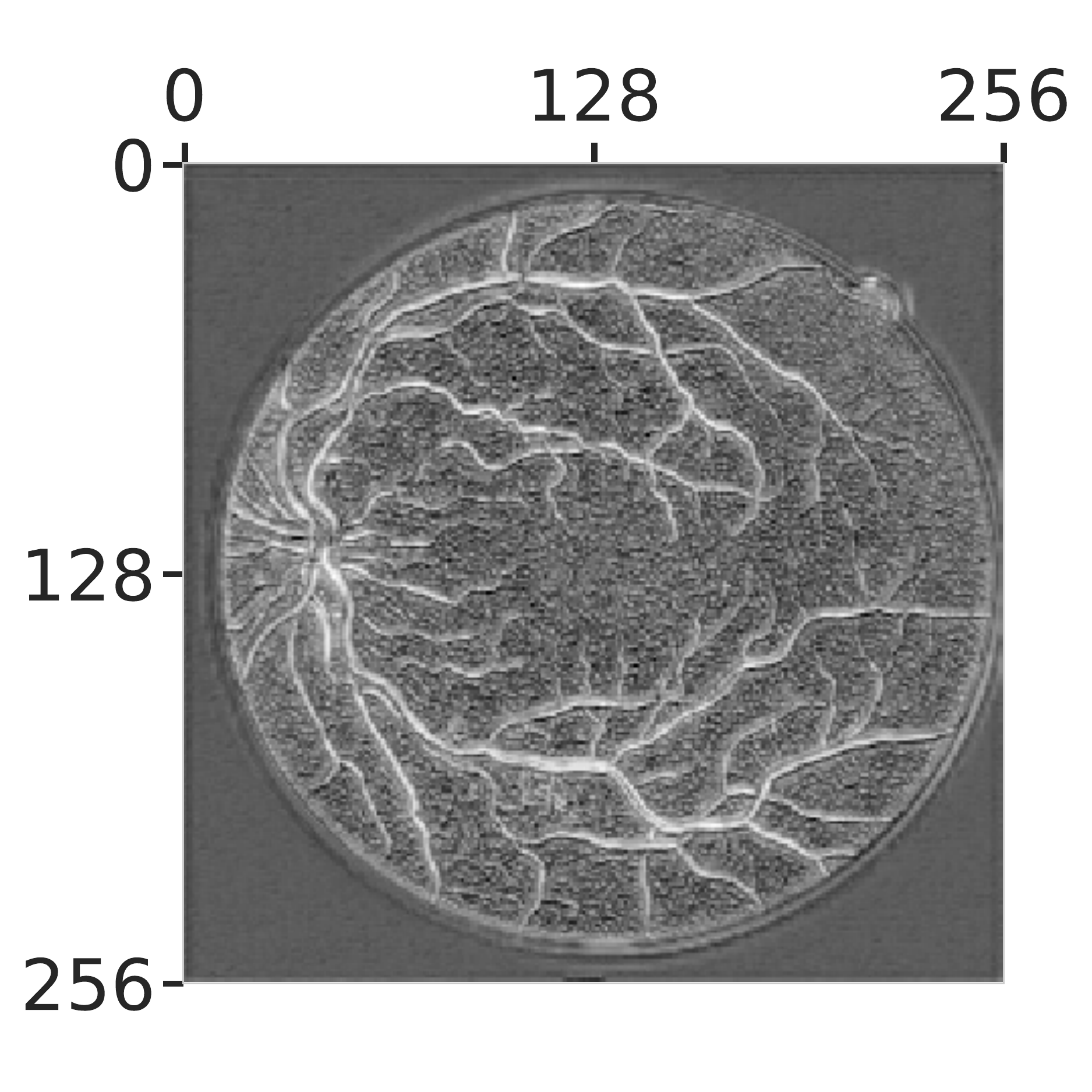}}%
    \subfloat[Encoder 2 (high)]{\includegraphics[width=.33\columnwidth]{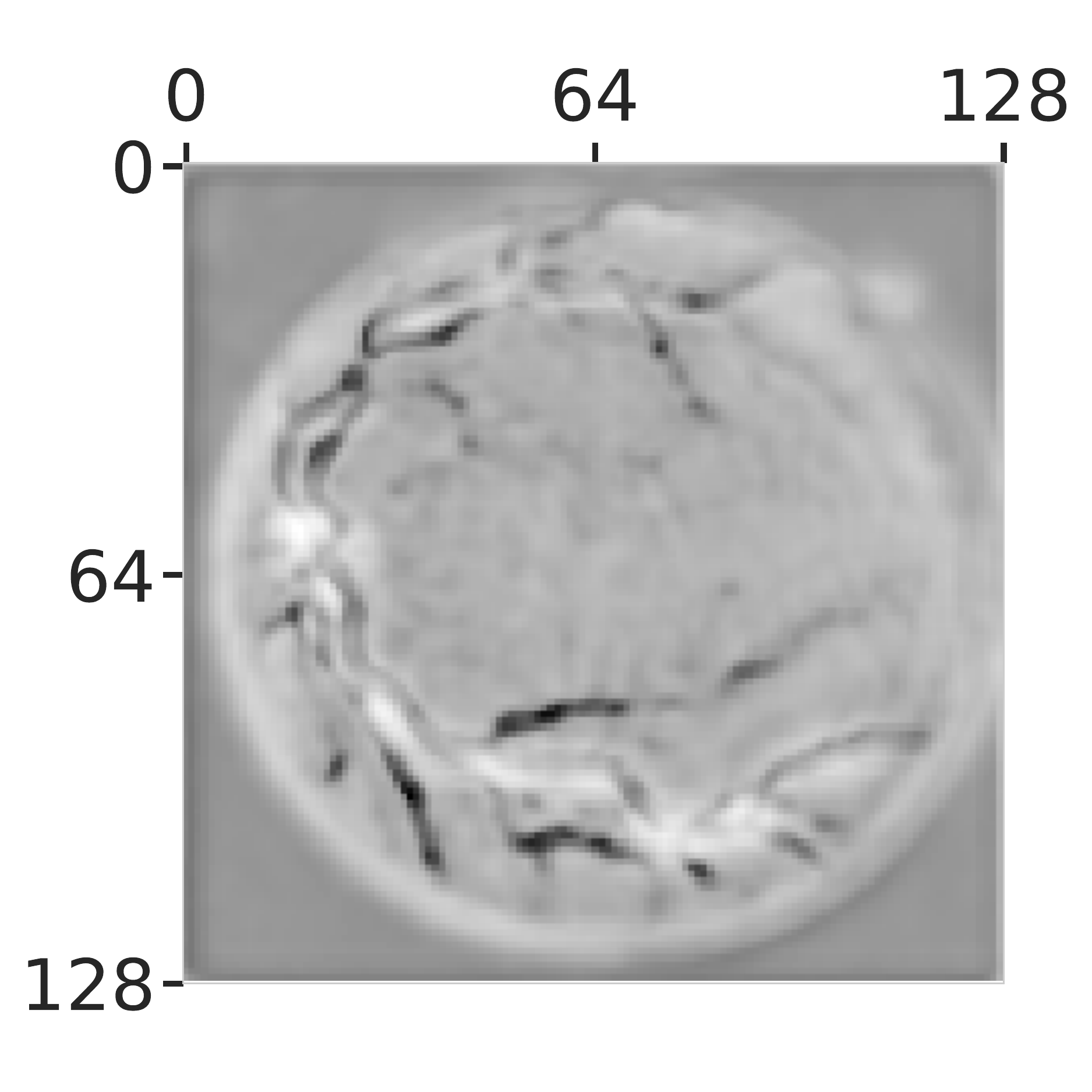}}%
    \subfloat[Decoder 2 (high)]{\includegraphics[width=.33\columnwidth]{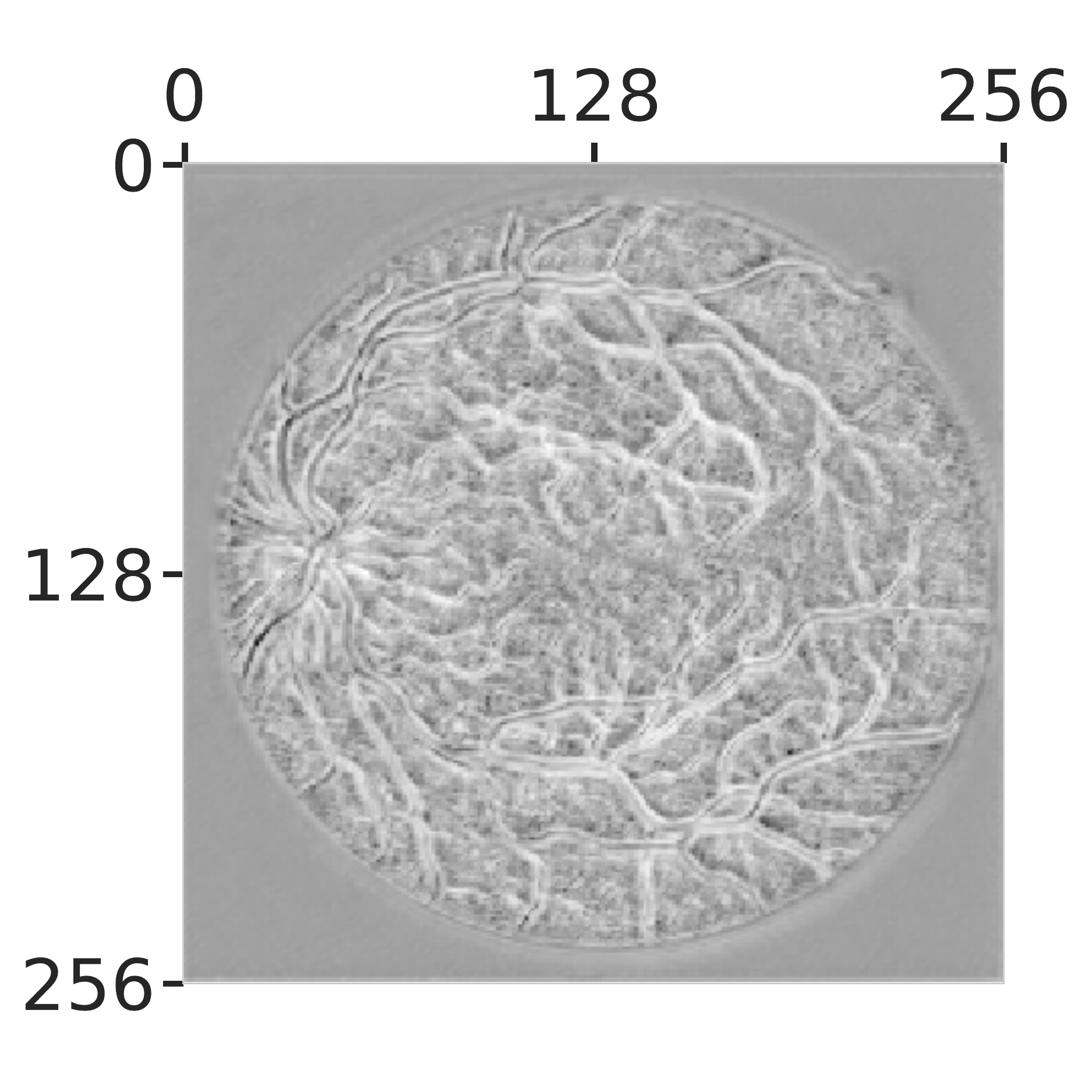}}%
    \subfloat[Decoder 1 (high)]{\includegraphics[width=.33\columnwidth]{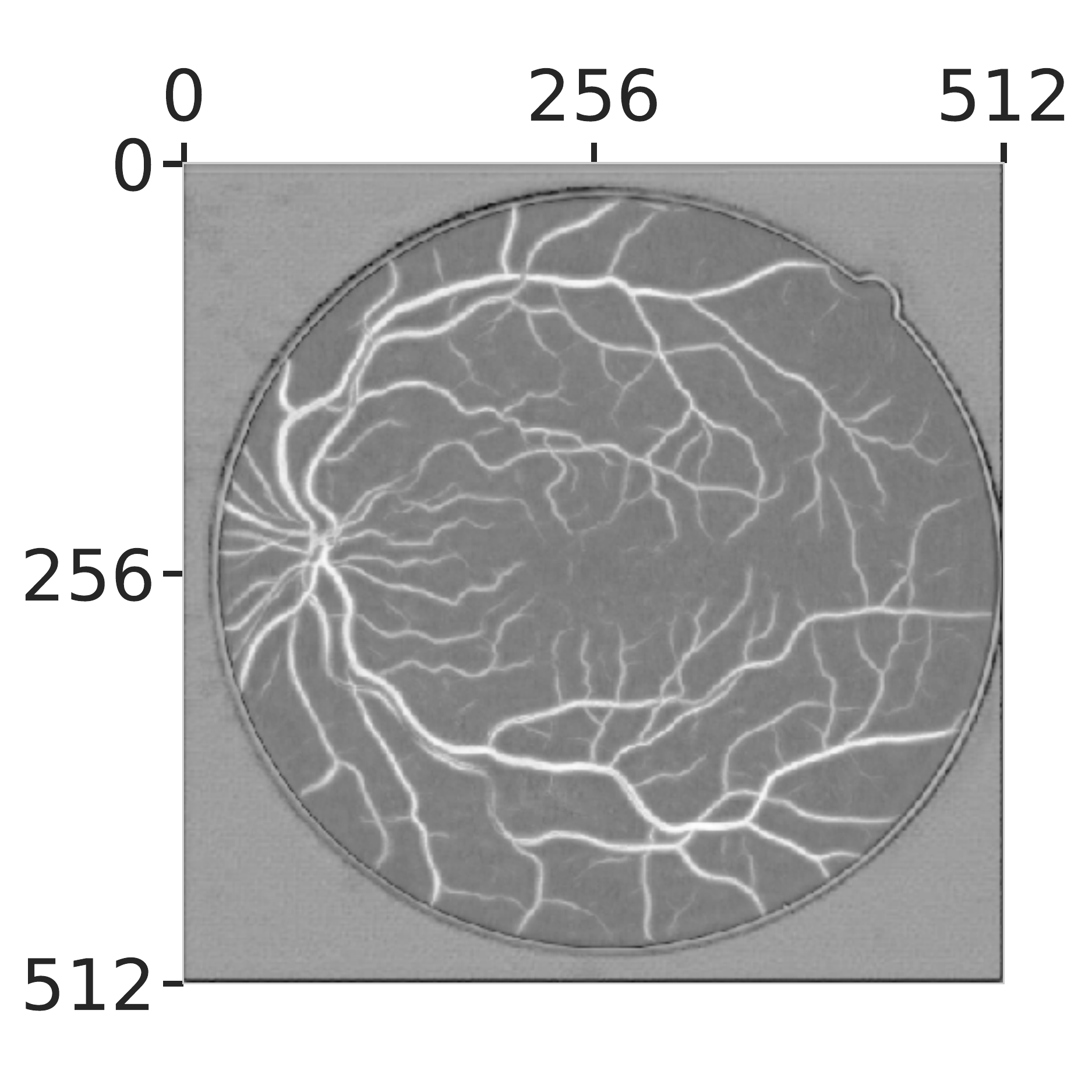}}%
    \subfloat[Decoder 0 (high)]{\includegraphics[width=.33\columnwidth]{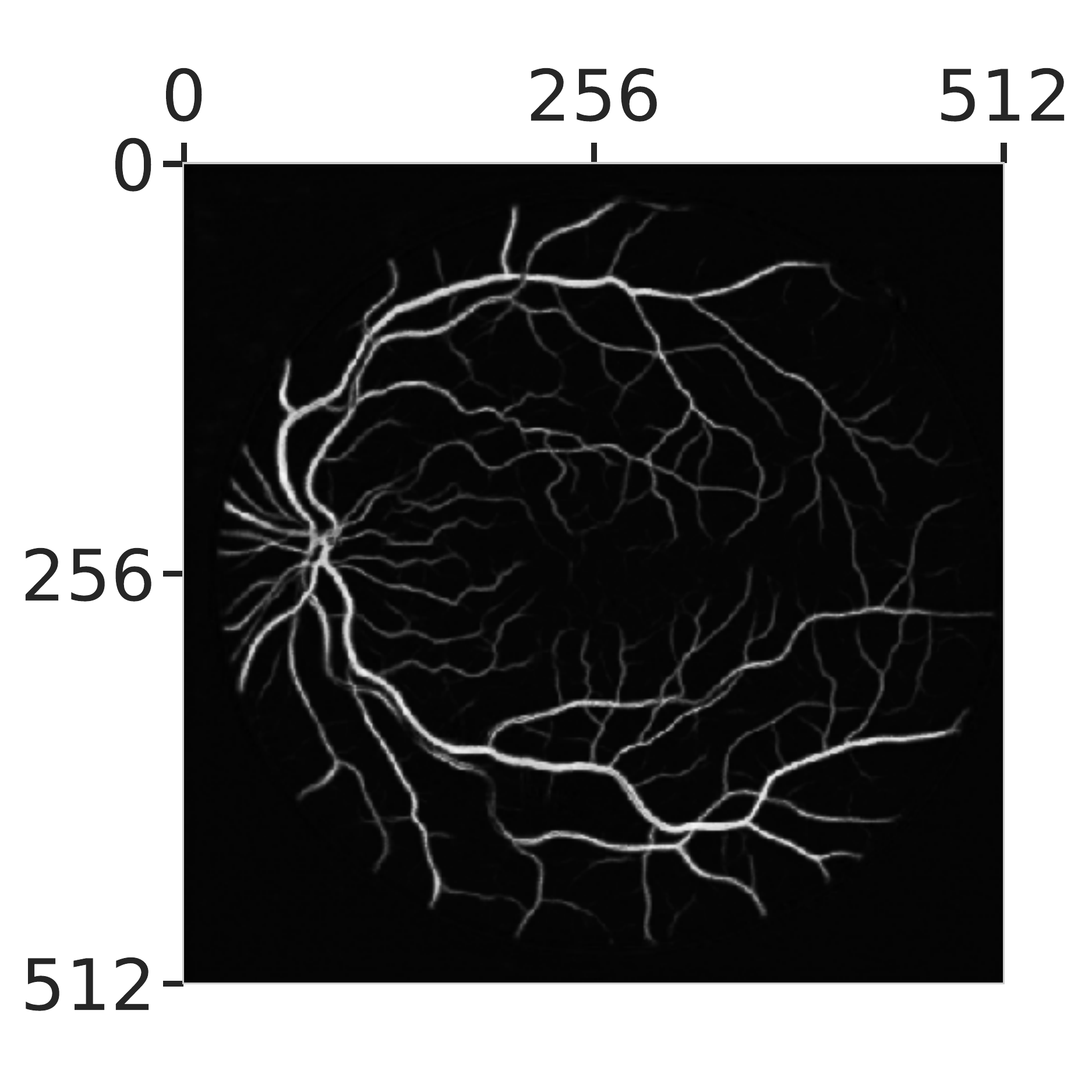}}
    \hfill\raggedright{}
    \subfloat[Encoder 0 (low)]{\includegraphics[width=.33\columnwidth]{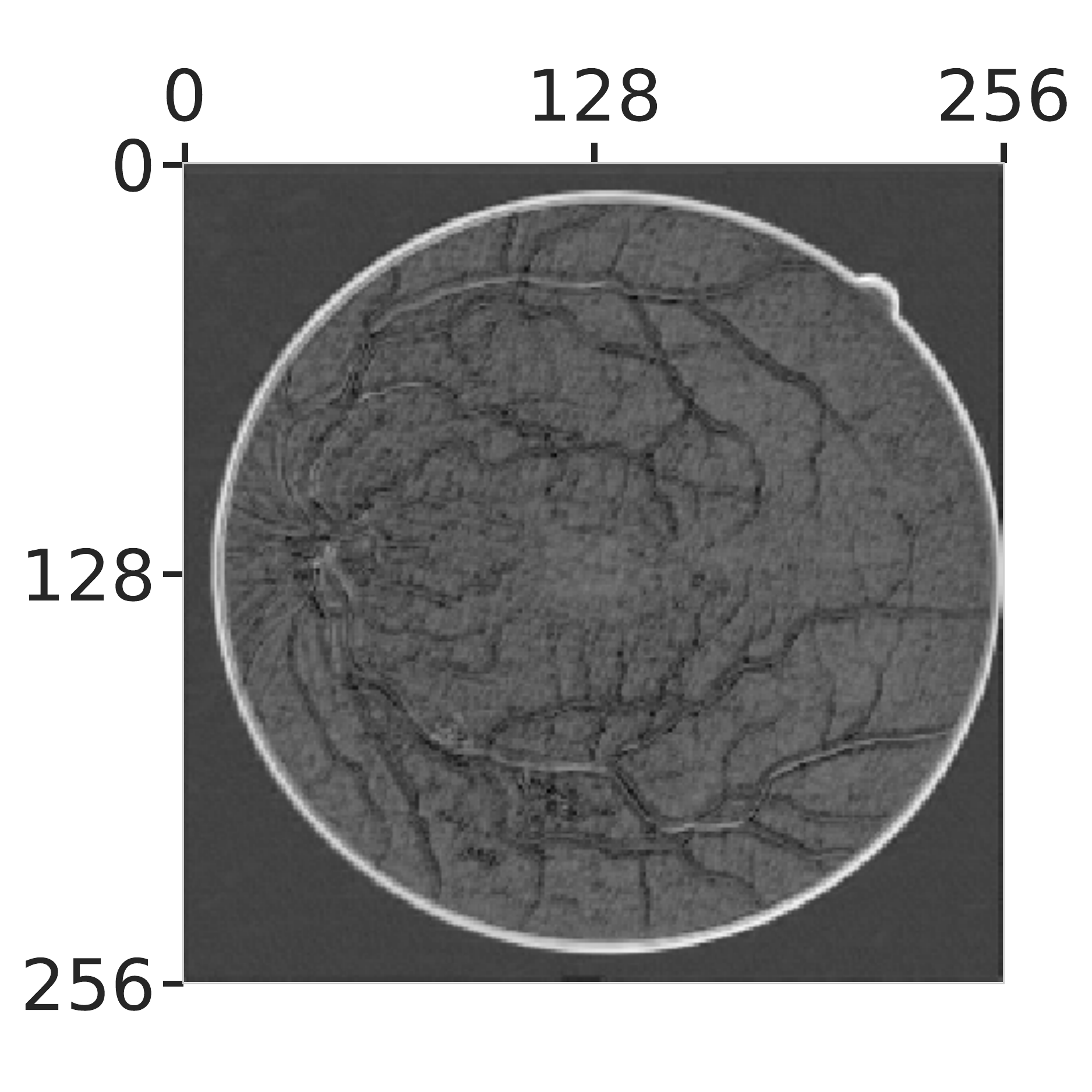}}%
    \subfloat[Encoder 1 (low)]{\includegraphics[width=.33\columnwidth]{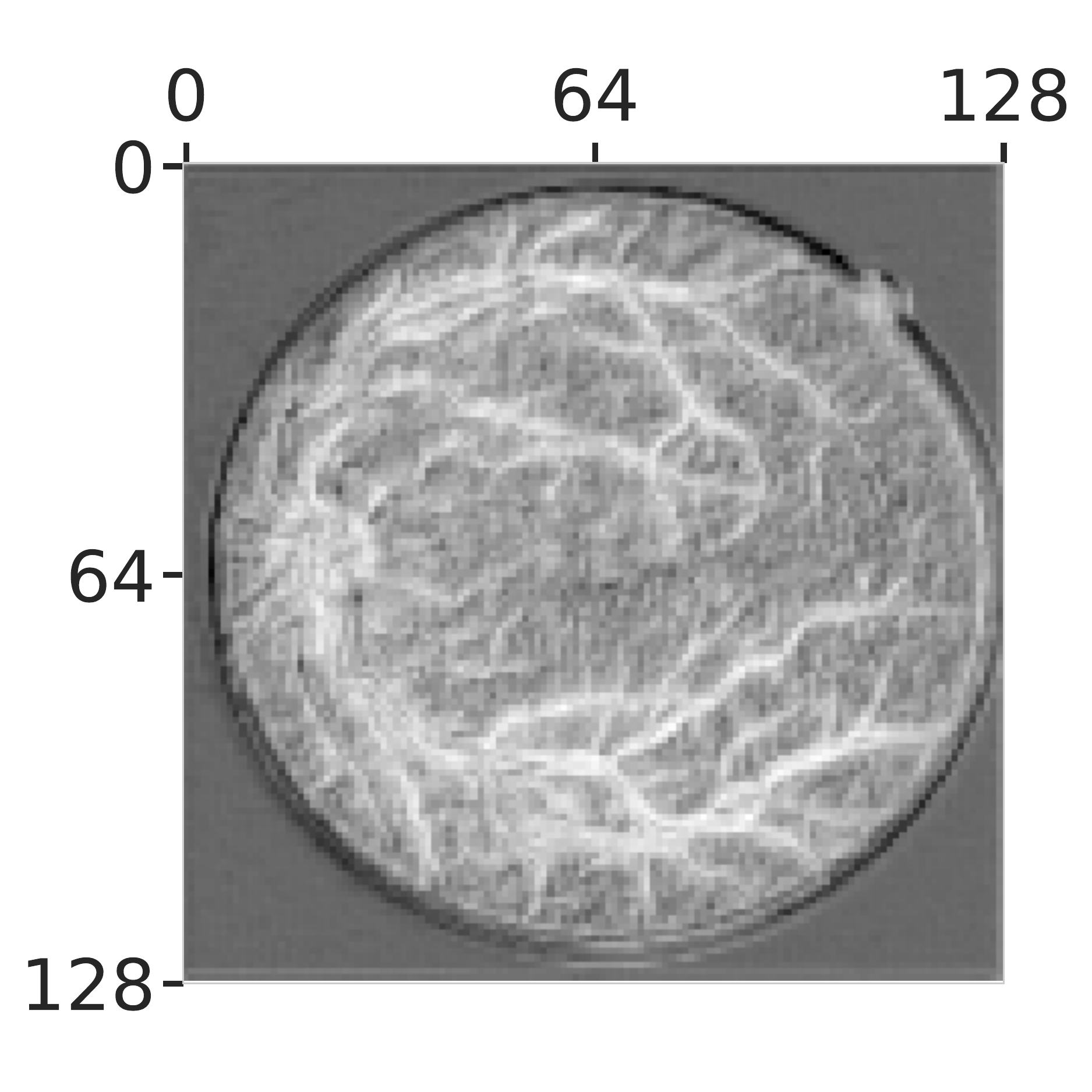}}%
    \subfloat[Encoder 2 (low)]{\includegraphics[width=.33\columnwidth]{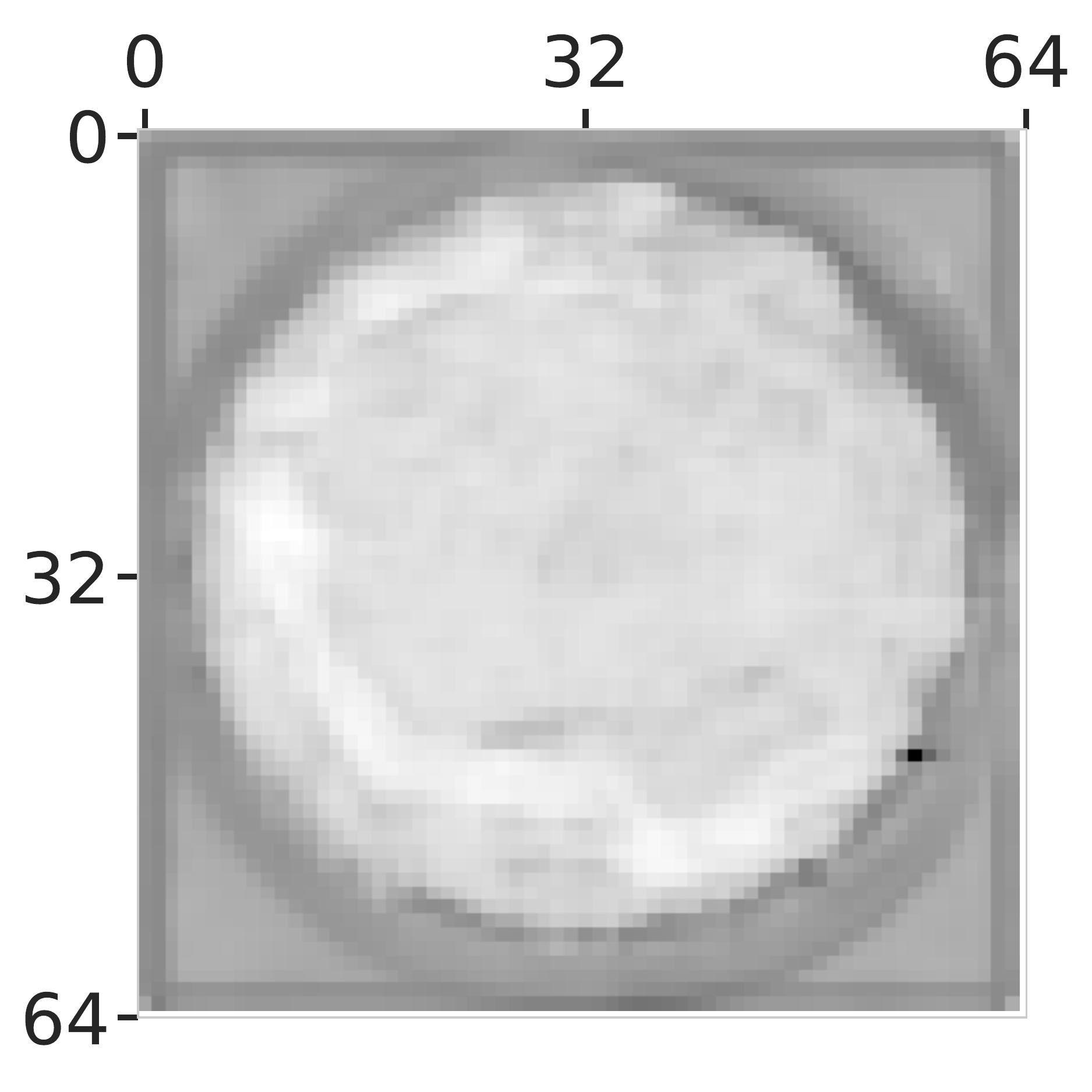}}%
    \subfloat[Decoder 2 (low)]{\includegraphics[width=.33\columnwidth]{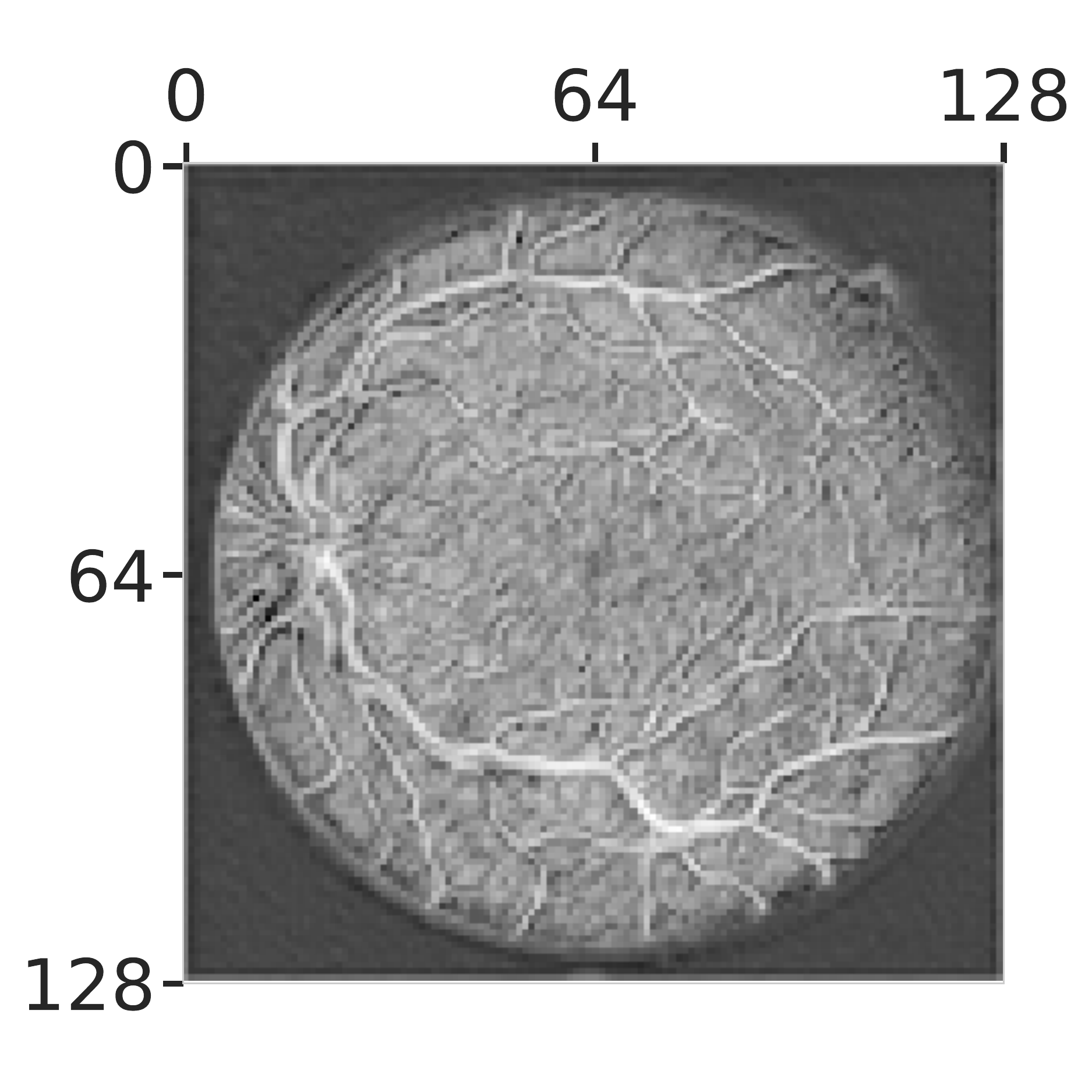}}%
    \subfloat[Decoder 1 (low)]{\includegraphics[width=.33\columnwidth]{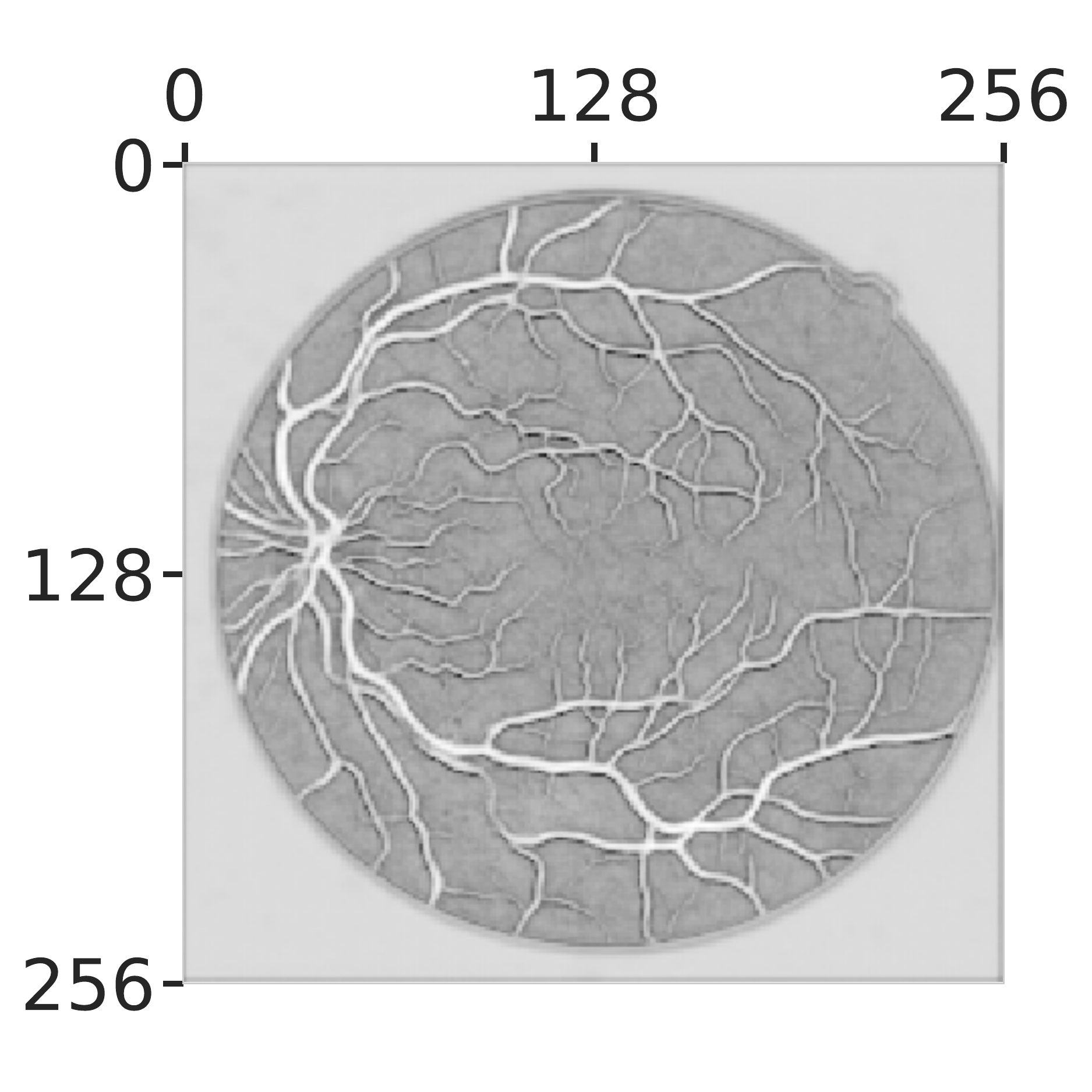}}
    \caption{High (first row) and low (second row) frequency kernel responses of encoders and decoders sampled at different abstract level.
        The first 3 columns show feature maps from different encoders,
        whereas the last 3 columns show the reconstructed vessel maps from different decoders.
        The height and width of the feature maps are noted on the axes.}\label{fig:feature}
\end{figure*}

In this section, a novel encoder-decoder based
neural network architecture named Octave UNet is proposed.
After end-to-end training, the proposed Octave UNet is
capable of extracting and decoding
hierarchical multifrequency features for segmenting retinal vasculature
in full-size fundus images.
The computation pipeline of the Octave UNet consists of two main processes,
i.e., feature encoding and decoding.
By utilizing octave convolutions and octave transposed convolutions,
we design multifrequency feature encoder blocks
and decoder blocks for hierarchical multifrequency feature
learning and decoding.
By stacking multiple encoder blocks sequentially
as shown in Fig.\ref{fig:OctaveUNet},
hierarchical multifrequency features can learn
to capture both the low frequency components
that describe the smoothly changing structures
such as the major vessels,
and high frequency components that describe
the abruptly changing details including
the fine details of thin vessels,
as shown in Fig.\ref{fig:feature}.

According to the feature encoding sequence shown in Fig.\ref{fig:OctaveUNet},
the feature maps lose spatial details and
the precise location information gradually,
while the spatial dimensions of the feature maps reduce step by step.
An example of this effect is shown in Fig.\ref{fig:feature}.
Only using features of high-abstract-level
that lack location information is insufficient
for generating precise segmentation results.
Inspired by the UNet\cite{Ronneberger:2015vw},
skip connections are adopted to concatenate
location-information-rich features
to the inputs of decoder blocks as shown in Fig.\ref{fig:OctaveUNet}.
The stacks of decoder blocks in combination with skip connections
shown in Fig.\ref{fig:OctaveUNet}
can facilitate the restoration of location information
and spatial details as illustrated
in the outputs of decoders in Fig.\ref{fig:feature}.

It is worth mentioning that the initial layer
of the Octave UNet shown in Fig.\ref{fig:OctaveUNet}
contains only computation of \(\hat{Y}^{H} = f^{H\rightarrow H}(X^{H})\)
and \(\hat{Y}^{L} = f^{H\rightarrow L}(X^{H})\),
where \(X^{H}\) is the input of fundus image
and the computation involving \(X^{L}\) is ignored.
Similarly, the final layer in Fig.\ref{fig:OctaveUNet} contains
only computation of \(\hat{Y}^{H} = f^{H\rightarrow H}(X^{H}) + f^{L\rightarrow H}(X^{L})\),
where \(\hat{Y}^{H}\) is the probability vessel map generated by the Octave UNet
and the computation for \(\hat{Y}^{L}\) is ignored.
For all the trainable kernels within octave convolution
and octave transposed convolution layers,
the kernel sizes are set to three, while unit strides
and one pixel zero paddings at both height and width are adopted.
Except for the final layer that is activated by sigmoid function
(\(\sigma(x) = \frac{1}{1+e^{-x}}\))
for performing dense binary classification,
ReLU activation (\(\sigma(x) = \max(x, 0)\))
is adopted for all the other layers.
Batch normalization\cite{Ioffe:2015ud} is also applied
after every layer with trainable parameters.
The Octave UNet can be trained in an end-to-end manner
on sample pairs of full-size fundus images and vessel ground truths.

\section{Datasets}\label{sec:Datasets}
The proposed method is evaluated on four publicly available
retinal fundus image datasets: DRIVE\cite{Staal:2004dd},
STARE\cite{Hoover:2000do}, CHASE\_DB1\cite{Fraz:2012gx},
and HRF\cite{Odstrcilik:2013ir}.
An overview of these four publicly available datasets
is provided in Table~\ref{tab:Datasets}.

\begin{table}[!htbp]
    \centering
    \caption{Overview of datasets adopted in this paper.}\label{tab:Datasets}
    \begin{tabular}{@{}llll@{}}
        \toprule
        Dataset    & Year & Description                    & Resolution          \\ \midrule
        DRIVE      & 2004 & \begin{tabular}[t]{@{}l@{}}40 in total,\\ 20 for training,\\ 20 for testing.\end{tabular} & \(565\times 584\)   \\
        STARE      & 2000 & \begin{tabular}[t]{@{}l@{}}20 in total,\\ 10 are abnormal.\end{tabular} & \(700\times 605\)   \\
        CHASE\_DB1 & 2011 & 28 in total.                   & \(999\times 960\)   \\
        HRF        & 2011 & \begin{tabular}[t]{@{}l@{}}45 in total,\\ 15 each for healthy,\\ diabetic and glaucoma.\end{tabular} & \(3504\times 2336\) \\ \bottomrule
    \end{tabular}
\end{table}

No post-processing steps are needed in the implementation.
However, to increase the limited diversity of fundus images,
a scheme of data augmentation pipeline is adopted,
which mainly includes steps such as horizontal or vertical flipping,
adjustment of brightness, saturation or contrast, and gamma adjustment.
Each of these steps has a trigger probability of \(50\% \).

\section{Experiments}\label{sec:Experiments}
\subsection{Evaluation metrics}\label{subsec:metrics}
Retinal vessel segmentation is often formulated as
a binary dense classification task,
i.e., predicting each pixel belonging to positive (vessel)
or negative (non-vessel) class within an input image.
As shown in Table~\ref{tab:Confusion},
a pixel prediction can fall into
one of the four categories,
i.e., True Positive (TP), True Negative (TP),
False Positive (FP), and False Negative (FN).
By plotting these pixels with different color,
e.g., TP with Green, FP with Red, TN with Black, and FN with Blue,
an analytical vessel map can be generated,
as shown in Fig.\ref{fig:best-and-worst-cases}.

\begin{table}[!htbp]
    \caption{A binary confusion matrix for vessel segmentation.}\label{tab:Confusion}
    \centering
    \begin{tabular}{@{}lll@{}}
        \toprule
        \multirow{2}{*}{Predicted class} & \multicolumn{2}{c}{Ground truth class}                       \\ \cmidrule(l){2-3}
                                         & Vessel                                 & Non-vessel          \\ \midrule
        Vessel                           & True Positive (TP)                     & False Negative (FN) \\
        Non-vessel                       & False Positive (FP)                    & True Negative (TP)  \\ \bottomrule
    \end{tabular}
\end{table}

As listed in Table~\ref{tab:metrics}, we adopt five commonly used metrics
for evaluation: accuracy (ACC), sensitivity (SE),
specificity (SP), F1 score (F1),
and area under Receiver Operating Characteristic curve (AUROC)
for comparison with state-of-the-art methods.
In addition, average precision (AP) is adopted to compare
the proposed method with the baseline method.

\begin{table}[!htbp]
    \centering
    \caption{Evaluation metrics adopted in this work.}\label{tab:metrics}
    \begin{tabular}{@{}ll@{}}
        \toprule
        Evaluation metric      & Description                                      \\ \midrule
        accuracy (ACC)         & \(\text{ACC = (TP + TN) / (TP + TN + FP + FN)}\) \\
        sensitivity (SE)       & \(\text{SE = TP / (TP + FN)}\)                   \\
        specificity (SP)       & \(\text{SP = TN / (TN + FP)}\)                   \\
        F1 score (F1)          & \(\text{F1 = (2 * TP) / (2 * TP + FP + FN)}\)    \\
        AUROC                  & Area under the ROC curve.                        \\
        average precision (AP) & Area under the precision-recall curve.           \\ \bottomrule
    \end{tabular}
\end{table}

For methods that use binary thresholding to obtain
the final segmentation results,
ACC, SE, SP and F1 are dependent on the binarization method.
In this paper, without special mentioning,
all threshold-sensitive metrics are calculated
by global thresholding with threshold \(\tau =0.5\).
On the other hand, calculating the area under ROC curve
requires first creating the ROC curve
by plotting the sensitivity
against the false positive rate (\(\text{FPR = FP / (TP + FN)}\))
at various threshold values.
Similarly, the average precision is calculated as the area under
the precision-recall curve which is plotted by
precision (\(\text{precision = TP / (TP + FP)}\)) against
recall (i.e., sensitivity) at various threshold values.

For a model that matches the ground truths perfectly with zero error,
its ACC, SE, SP, F1, AUROC and AP should all hit the best score: one.

\subsection{Experiment settings}\label{subsec:Expsetup}
\subsubsection{Loss function}\label{subsubsec:Loss}
To alleviate the effect of imbalanced-classes problem
(i.e., the vessel pixel to non-vessel pixel ratio is about \(1/9\)),
class weighted binary cross-entropy as shown in (\ref{eq:loss})
is adopted as the loss function for training,
where the positive class weight \(w_{pos} = p_{pos} / p_{neg}\)
is calculated as the ratio of the positive pixel count \(p_{pos}\)
to the negative pixel count \(p_{neg}\) of the training set.
Equation (\ref{eq:loss}) measures the loss of a batch of \(m\) samples,
in which \(y_{n}\) and \(\hat{y}_{n}\) denote
the ground truth and model prediction of \(n\text{-th}\) sample, respectively.

\setlength{\arraycolsep}{0.0em}
\begin{eqnarray}
    \label{eq:loss}
    L &{=}& -\sum_{n=1}^{m}{(w_{pos} y_{n} \log{(\hat{y}_{n})} + (1-y_{n}) \log{(1-\hat{y}_{n})})}
\end{eqnarray}
\setlength{\arraycolsep}{5pt}

\subsubsection{Training details}\label{subsubsec:Train}
All models are trained from scratch with Adam optimizer\cite{Kingma:2015us}
with default hyper-parameters (e.g., \(\beta_1 = 0.9\) and \(\beta_2 = 0.999\)).
The initial learning rate is set to \(\eta=0.001\).
A shrinking schedule is applied
for the current learning rate \(\eta_{i}\) as \(\eta_{i} = 0.9\eta_{i-1}\)
after the value of loss function has been saturated for 10 epochs.
The training process runs for a total of 1000 epochs
to ensure convergence of the model.
All trainable kernels are initialized with He initialization\cite{He:2015dj},
and no pre-trained model parameters are used.

\subsubsection{Splitting of training and testing set}\label{subsubsec:Split}
Except for the DRIVE dataset which has a conventional splitting of
training and testing set,
the strategy of leave-one-out validation is adopted for
STARE, CHASE\_DB1, and HRF datasets.
Specifically, all images are tested using a model
trained on the other images within the same dataset.
This strategy for generating training and testing set is also adopted by
recent works in\cite{Li:2016jl,Liskowski:2016,Staal:2004dd,Mo:2017io}.
Only the results on test samples are reported.

\subsection{Comparison with baseline model}\label{subsec:ablation}
\begin{figure}[!htbp]
    \centering
    \includegraphics[width=\columnwidth]{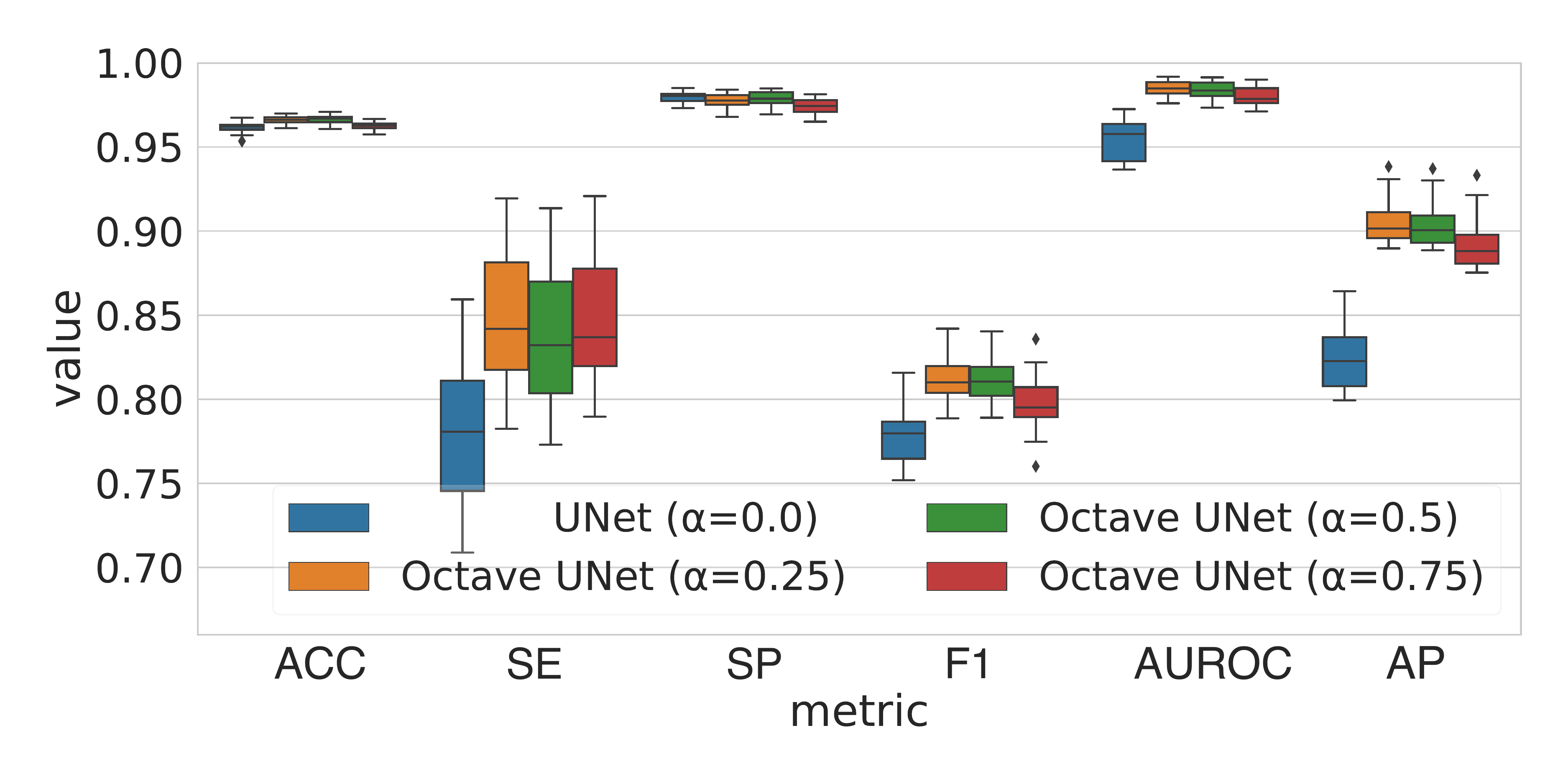}
    \caption{Box plot comparing the baseline UNet and the Octave UNets on the DRIVE dataset.}\label{fig:boxplot}
\end{figure}

\begin{figure*}[!htbp]
    \centering
    \includegraphics[width=\textwidth]{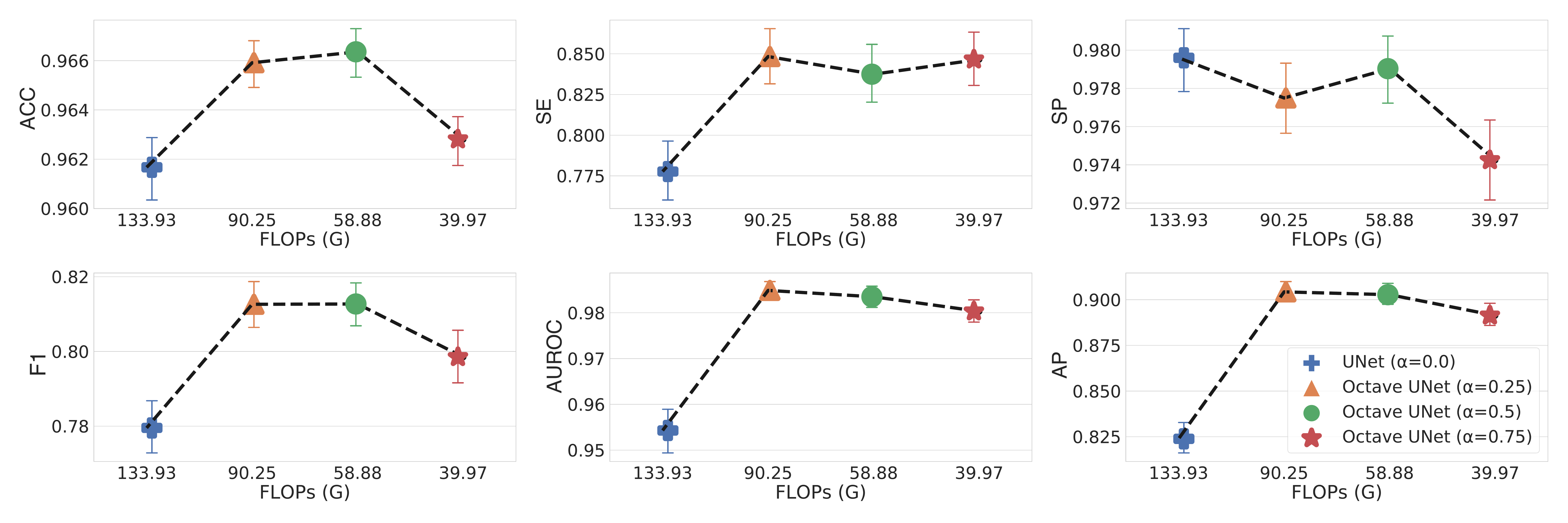}
    \caption{Comparison of performance and computational expenditure of the baseline UNet and the Octave UNets on the DRIVE dataset.}\label{fig:pareto}
\end{figure*}

To compare the proposed Octave UNet with the baseline UNet\cite{Ronneberger:2015vw},
a controlled experiment is conducted,
in which all factors are held constant except
for the hyper-parameter \(\alpha \)
that affects the ratio of number of channels of low-frequency features
to the total number of channels.
(e.g., the baseline UNet has a hyper-parameter of \(\alpha = 0\)).
The experimental results of the baseline model and the Octave UNet models
with different \(\alpha \in [0.25, 0.5, 0.75]\)
are reported in Table~\ref{tab:ablation}
and presented in Fig.\ref{fig:boxplot},
which demonstrates that the multi-frequency feature learning indeed improves
the performance of the baseline model in terms of
all metrics except for specificity, which only reduces very slightly.
Furthermore, the performance against computational expenditure curves
in Fig.\ref{fig:pareto} shows that the Octave UNet with \(\alpha = 0.5\)
is at a preferable selection, which outperforms
the baseline UNet with a reduction of
about half of the computational expenditure.

\begin{table*}[!htbp]
    \centering
    \caption{Performance comparison of the baseline UNet and the Octave UNets on the DRIVE dataset.}\label{tab:ablation}
    \begin{tabular}{@{}lllllllll@{}}
        \toprule
        Model                         & \# Parameter (M) & FLOPs (G) & ACC             & SE              & SP              & F1              & AUROC           & AP              \\ \midrule
        baseline UNet (\(\alpha=0\))  & 16.53            & 133.93    & 0.9616          & 0.7768          & \textbf{0.9796} & 0.7795          & 0.9543          & 0.8238          \\
        Octave UNet (\(\alpha=0.25\)) & 16.54            & 90.25     & 0.9659          & \textbf{0.8483} & 0.9775          & 0.8126          & \textbf{0.9849} & \textbf{0.9042} \\
        Octave UNet (\(\alpha=0.5\))  & 16.54            & 58.88     & \textbf{0.9663} & 0.8375          & 0.9790          & \textbf{0.8127} & 0.9835          & 0.9027          \\
        Octave UNet (\(\alpha=0.75\)) & 16.54            & 39.97     & 0.9628          & 0.8465          & 0.9742          & 0.7985          & 0.9803          & 0.8913          \\ \bottomrule
    \end{tabular}
\end{table*}

\begin{figure}[!htbp]
    \centering
    \captionsetup{width=.4\columnwidth}
    \subfloat[Fundus image]{\includegraphics[width=.49\columnwidth]{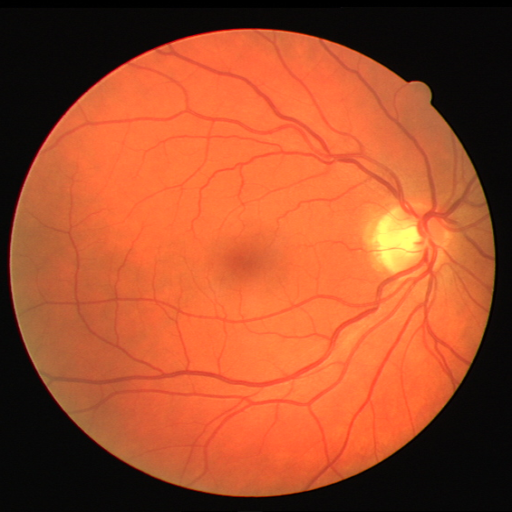}}\label{fig:case-images}%
    \subfloat[Ground truth]{\includegraphics[width=.49\columnwidth]{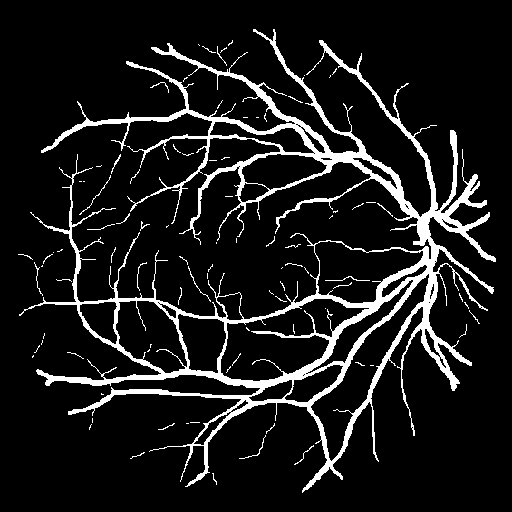}}\label{fig:case-targets}
    \subfloat[Probability map of the baseline UNet]{\includegraphics[width=.49\columnwidth]{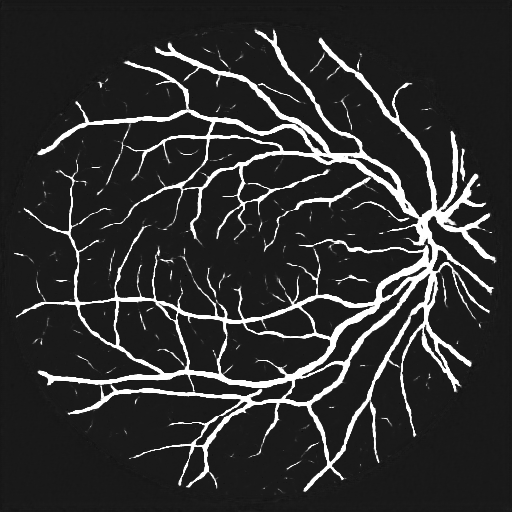}}\label{fig:case-unet}%
    \subfloat[Probability map of the Octave UNet]{\includegraphics[width=.49\columnwidth]{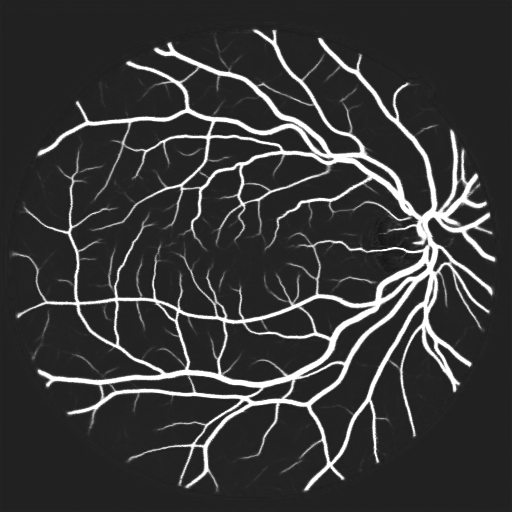}}\label{fig:case-octave}
    \caption{An example in the DRIVE dataset, with (a) the original fundus image, (b) the corresponding ground truth, (c) the probability map generated by the baseline UNet, and (d) the probability map generated by the Octave UNet.
        The Octave UNet shows better performance in capturing the thin vessels while having better connectivity.}\label{fig:case-study}
\end{figure}

An example of the probability maps generated by
the baseline UNet and the Octave UNet
are presented in Fig.\ref{fig:case-study},
which shows that the Octave UNet
can better capture the fine details of thin vessels
and segment the major vessels with better connectivity
than the baseline UNet.

\begin{figure}[!htbp]
    \centering
    \subfloat[baseline UNet]{\includegraphics[width=.33\columnwidth]{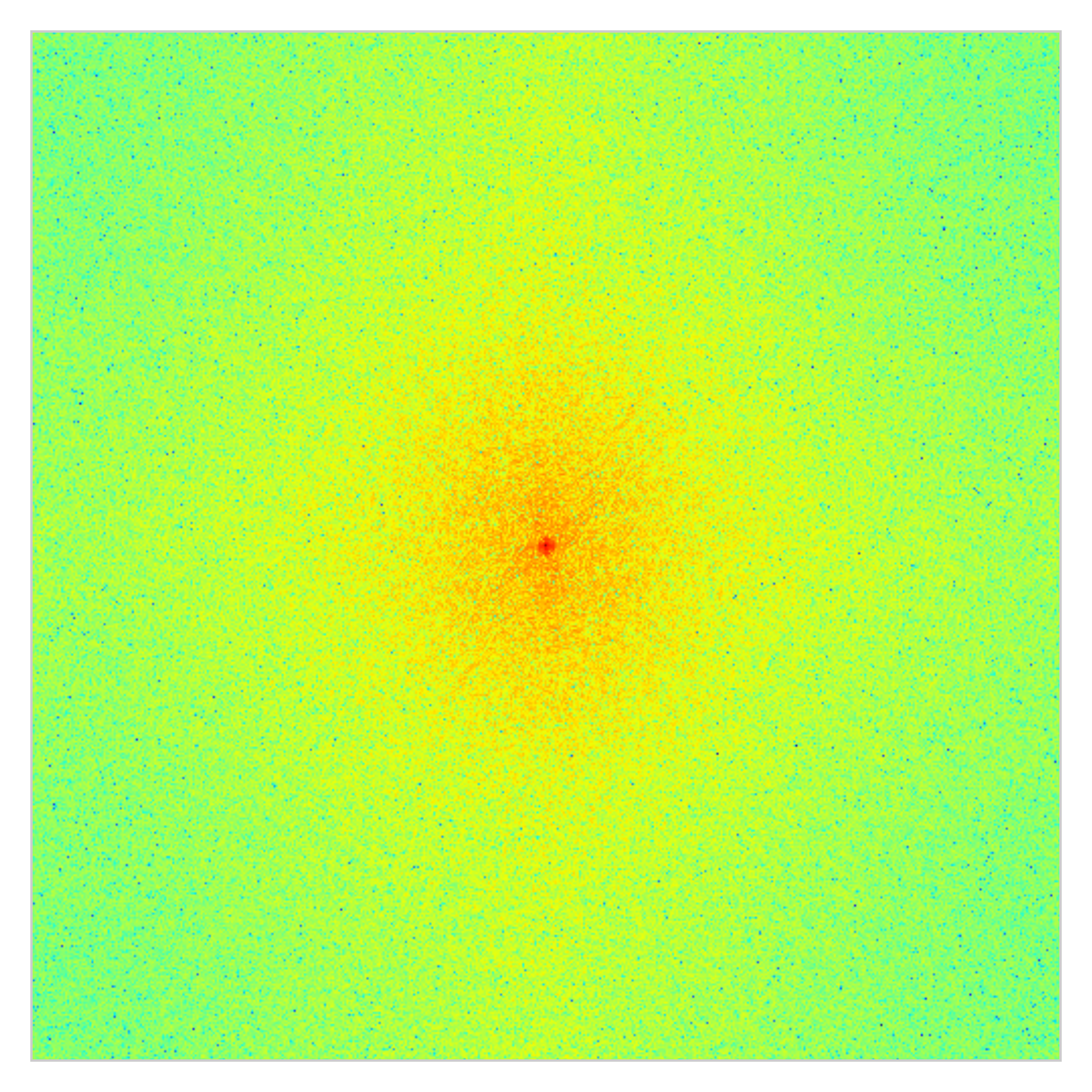}}\label{fig:feature-analysis-unet}%
    \subfloat[Octave UNet (high)]{\includegraphics[width=.33\columnwidth]{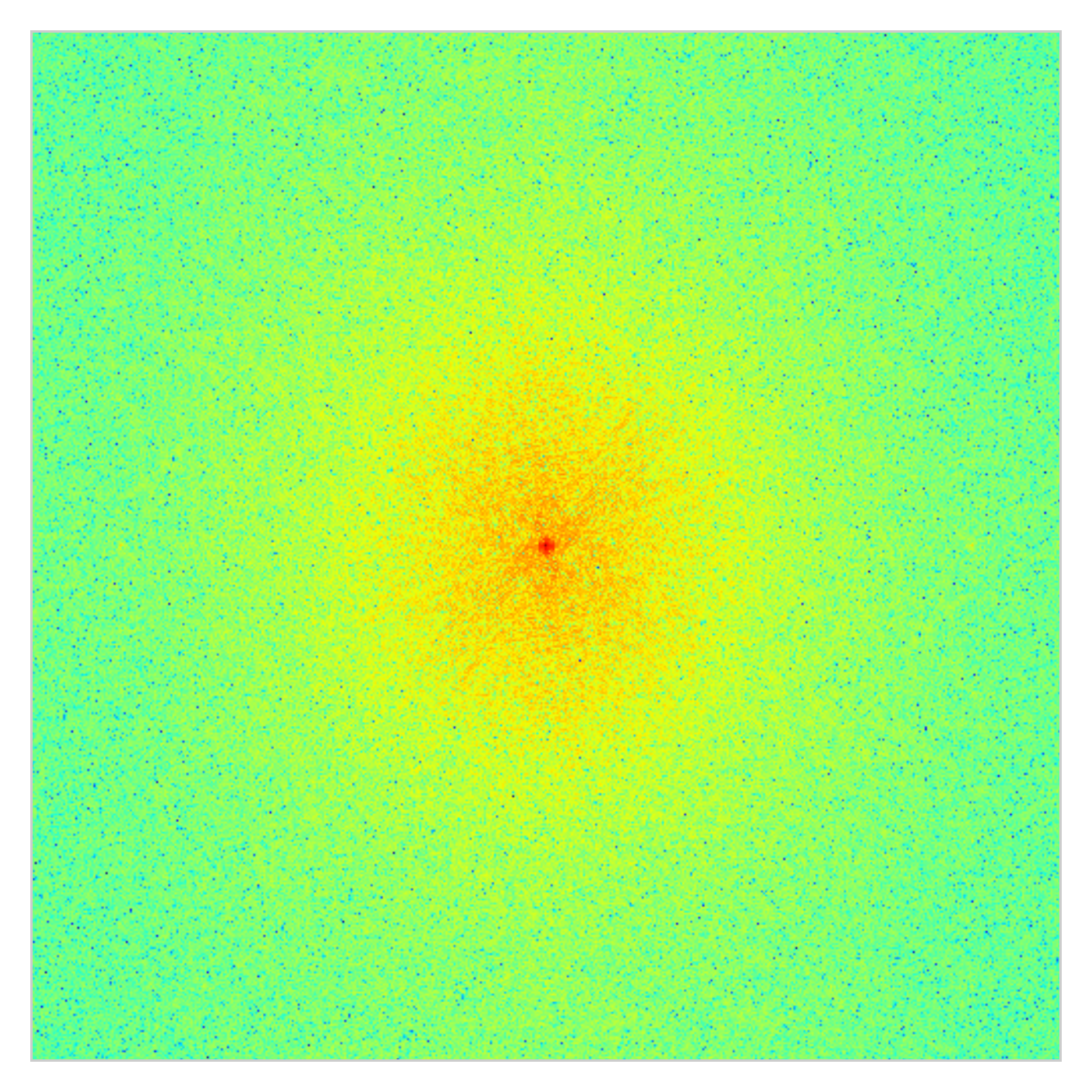}}\label{fig:feature-analysis-octave-high}%
    \subfloat[Octave UNet (low)]{\includegraphics[width=.33\columnwidth]{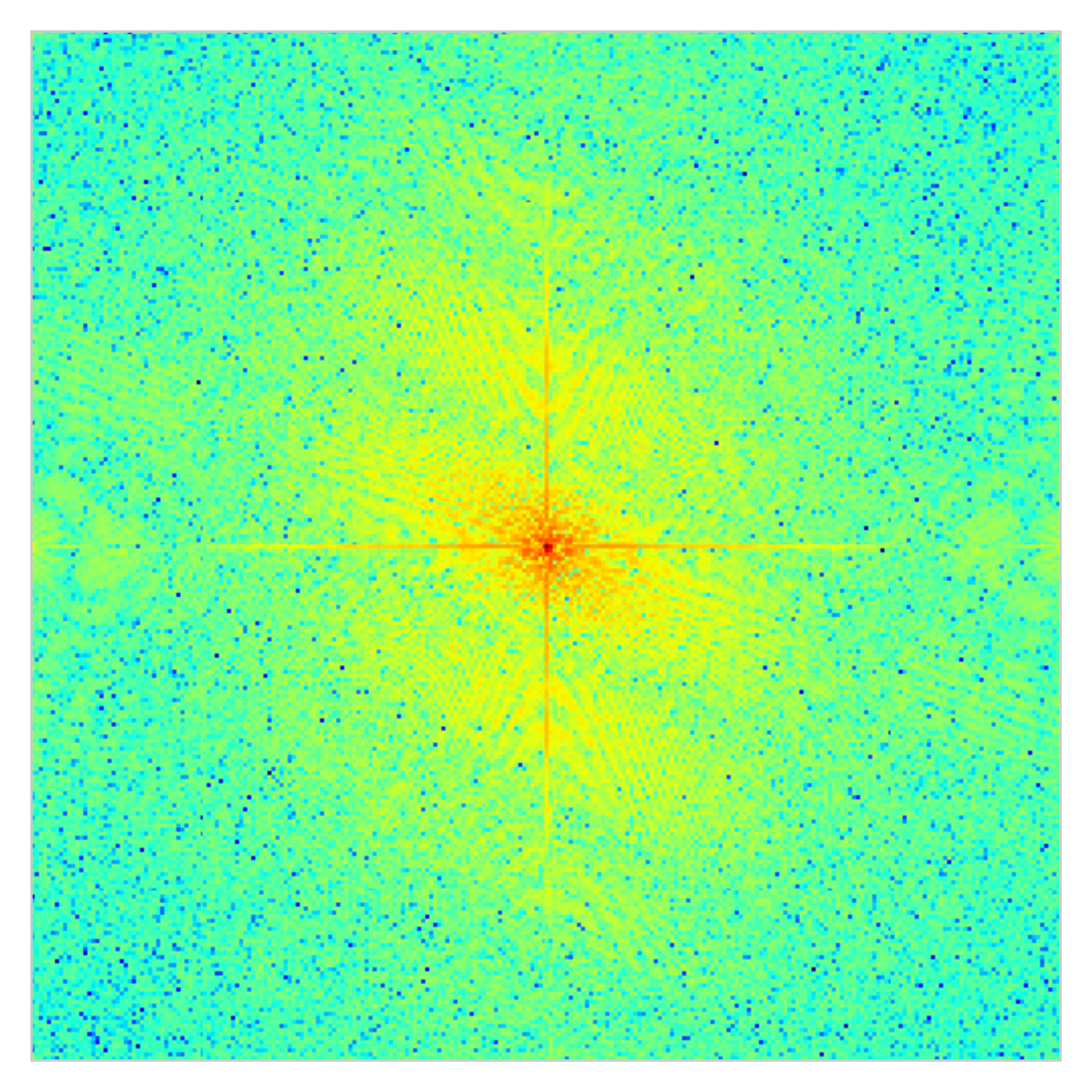}}\label{fig:feature-analysis-octave-low}
    \subfloat[Frequency spectrums]{\includegraphics[width=\columnwidth]{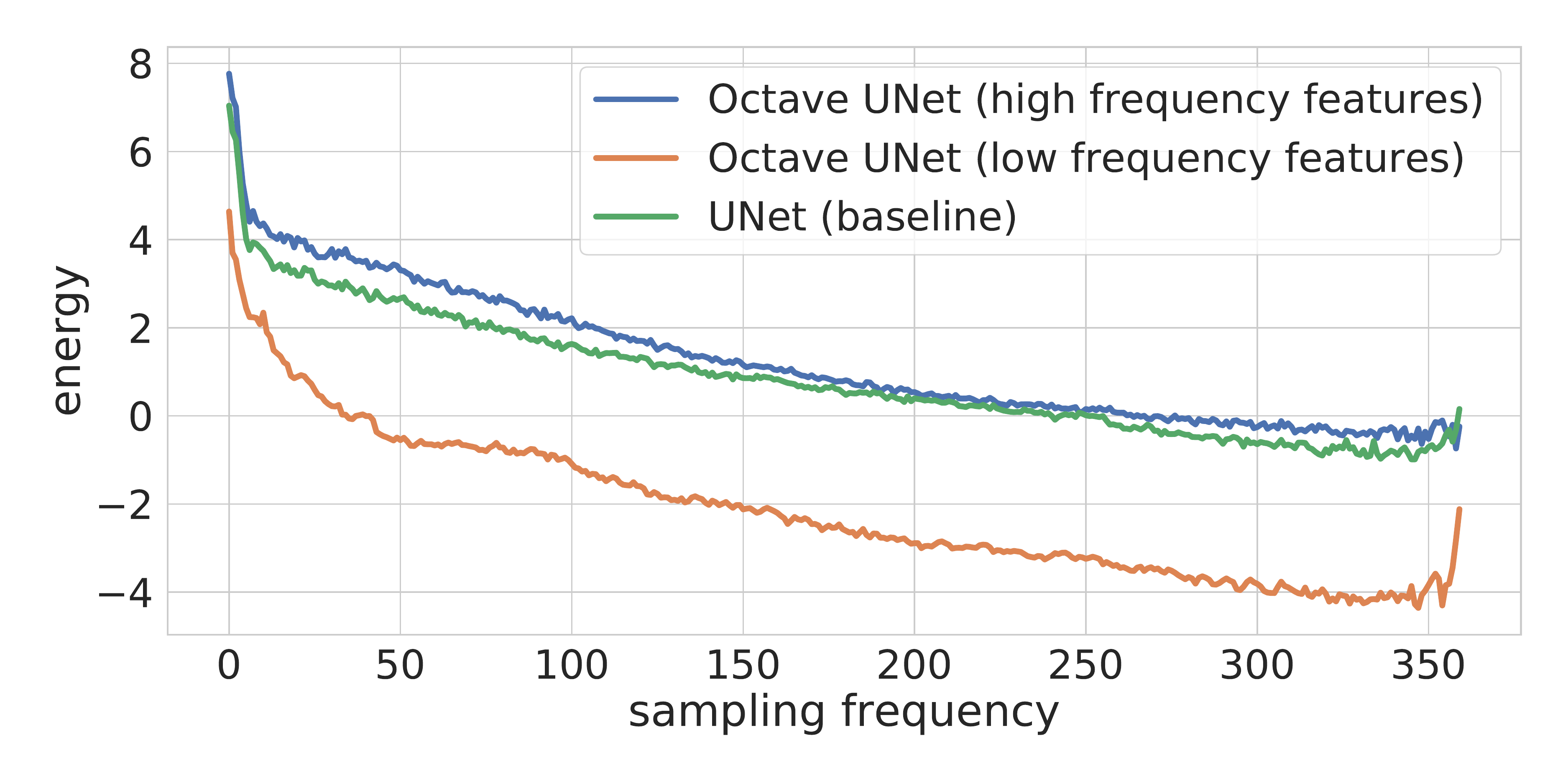}}\label{fig:feature-analysis-spectrums}
    \caption{Frequency analysis of feature maps.
        The two dimensional energy maps (a), (b) and (c) are the responses of Fourier transformation of feature maps obtained from the baseline UNet,
        high frequency features of the Octave UNet and low frequency features of the Octave UNet, respectively.
        The one dimensional spectrum curves in (d) are generated from the two dimensional energy maps, respectively.
    }\label{fig:feature-analysis}
\end{figure}

Moreover, a frequency analysis of the feature maps learned by the
Octave UNet and the baseline UNet is conducted and
the results are presented in Fig.\ref{fig:feature-analysis}.
Specifically, the feature maps generated by
the intermediate encoders and decoders of
the Octave UNet and the baseline UNet
are first obtained by feeding both the models with
the same fundus images in DRIVE dataset.
Then the two dimensional Fast Fourier Transformation (FFT) is applied
on each feature map to obtain the energy map of frequencies,
followed by shifting the low frequency components of the energy map
to the center while the high frequency components to the edges.
Finally, the two dimensional energy maps of shifted FFT response
generated from the baseline UNet, the high frequency group of the Octave UNet
and the low frequency group of the Octave UNet are respectively averaged
to generate the final results shown
in (a), (b), and (c) of Fig.\ref{fig:feature-analysis}.
Furthermore, to make it easier to distinguish the difference,
the final results of two dimensional energy maps
are transformed into one dimensional signals
as shown in (d) of Fig.\ref{fig:feature-analysis}
by first averaging pixels representing the same frequencies
and then sorting the averaged pixel values
according to the frequencies their represents.

It can be observed that the features learned by
the Octave UNet obtain different frequency characteristics
than those of the baseline UNet.
Specifically, the energy of the low frequency group of the Octave UNet is
more focused around the zero frequency than
those of the high frequency group,
whereas the energy of the high frequency group of the Octave UNet
contains more high frequency components than
the those of the baseline UNet.
Combining with the experimental results,
this observation supports the hypothesis that
adopting multi-frequency feature learning indeed improves
the performance of segmenting retinal vessels with varying sizes and shapes.

\subsection{Comparison with other state-of-the-art methods}\label{subsec:Compare}
The performance comparison of the proposed method
and the state-of-the-art methods are reported
in Table~\ref{tab:DRIVE}, Table~\ref{tab:STARE},
Table~\ref{tab:CHASE}, and Table~\ref{tab:CHASE}
for DRIVE, STARE, CHASE\_DB1, and HRF dataset, respectively.
The proposed method outperforms all the other
state-of-the-art methods in terms of ACC, SE, and F1 on all datasets.
Specifically, the proposed method achieves the best
performance in all metrics on CHASE\_DB1 and HRF datasets.
For DRIVE dataset, the proposed method
achieves best ACC, SE, F1, and AUROC,
while its SP is slightly lower
than the patch based R2-UNet\cite{Alom:2019id}.
For STARE dataset, the proposed Octave UNet obtains
metrics of SP and AUROC comparable to
the two patch based methods:
DeepVessel\cite{Liskowski:2016} and R2-UNet\cite{Alom:2019id},
while achieving best performance on all the other metrics.
Overall, the proposed method achieves better or comparable performance
against the other state-of-the-art methods.

\begin{table}[!htbp]
    \centering
    \caption{Comparison with other state-of-the-art methods on the DRIVE dataset.}\label{tab:DRIVE}
    \resizebox{\columnwidth}{!}{
        \begin{tabular}{llllllll}
            \toprule
            Methods                                                & Year & ACC             & SE              & SP              & F1              & AUROC           \\ \midrule
            Unsupervised Methods                                   &      &                 &                 &                 &                 &                 \\ \midrule
            Zena and Klein\cite{Zana:2001dw}                       & 2001 & 0.9377          & 0.6971          & 0.9769          & N/A             & 0.8984          \\
            Mendonca and Campilho\cite{Mendonca:2006dq}            & 2006 & 0.9452          & 0.7344          & 0.9764          & N/A             & N/A             \\
            Al-Diri \textit{et al.}\cite{AlDiri:2009jh}            & 2009 & 0.9258          & 0.7282          & 0.9551          & N/A             & N/A             \\
            Miri and Mahloojifar\cite{Miri:2011dn}                 & 2010 & 0.9458          & 0.7352          & 0.9795          & N/A             & N/A             \\
            You \textit{et al.}\cite{You:2011do}                   & 2011 & 0.9434          & 0.7410          & 0.9751          & N/A             & N/A             \\
            Fraz \textit{et al.}\cite{Fraz:2012ef}                 & 2012 & 0.9430          & 0.7152          & 0.9768          & N/A             & N/A             \\
            Fathi \textit{et al.}\cite{Fathi:2013di}               & 2013 & N/A             & 0.7768          & 0.9759          & 0.7669          & N/A             \\
            Roychowdhury \textit{et al.}\cite{Roychowdhury:2015da} & 2015 & 0.9494          & 0.7395          & 0.9782          & N/A             & N/A             \\
            Fan \textit{et al.}\cite{Fan:2019dn}                   & 2019 & 0.9600          & 0.7360          & 0.9810          & N/A             & N/A             \\ \midrule
            Supervised Methods                                     &      &                 &                 &                 &                 &                 \\ \midrule
            Staal \textit{et al.}\cite{Staal:2004dd}               & 2004 & 0.9441          & 0.7194          & 0.9773          & N/A             & 0.9520          \\
            Marin \textit{et al.}\cite{Marino:2006jf}              & 2011 & 0.9452          & 0.7067          & 0.9801          & N/A             & 0.9588          \\
            Fraz \textit{et al.}\cite{Fraz:2012gx}                 & 2012 & 0.9480          & 0.7460          & 0.9807          & N/A             & 0.9747          \\
            Cheng \textit{et al.}\cite{Cheng:2014hg}               & 2014 & 0.9472          & 0.7252          & 0.9778          & N/A             & 0.9648          \\
            Vega \textit{et al.}\cite{Vega:2015kh}                 & 2015 & 0.9412          & 0.7444          & 0.9612          & 0.6884          & N/A             \\
            Antiga\cite{Antiga:2016wf}                             & 2016 & 0.9548          & 0.7642          & 0.9826          & 0.8115          & 0.9775          \\
            Fan \textit{et al.}\cite{Fan:2016kd}                   & 2016 & 0.9614          & 0.7191          & 0.9849          & N/A             & N/A             \\
            Fan and Mo\cite{Fan:2016ie}                            & 2016 & 0.9612          & 0.7814          & 0.9788          & N/A             & N/A             \\
            Liskowski \textit{et al.}\cite{Liskowski:2016}         & 2016 & 0.9535          & 0.7811          & 0.9807          & N/A             & 0.9790          \\
            Li \textit{et al.}\cite{Li:2016jl}                     & 2016 & 0.9527          & 0.7569          & 0.9816          & N/A             & 0.9738          \\
            Orlando \textit{et al.}\cite{Orlando:2017iy}           & 2017 & N/A             & 0.7897          & 0.9684          & 0.7857          & N/A             \\
            Mo and Zhang\cite{Mo:2017io}                           & 2017 & 0.9521          & 0.7779          & 0.9780          & N/A             & 0.9782          \\
            Xiao \textit{et al.}\cite{Xiao:2018hs}                 & 2018 & 0.9655          & 0.7715          & N/A             & N/A             & N/A             \\
            Alom \textit{et al.}\cite{Alom:2019id}                 & 2019 & 0.9556          & 0.7792          & \textbf{0.9813} & 0.8117          & 0.9784          \\
            Lei \textit{et al.}\cite{Lei:2020}                     & 2019 & 0.9607          & 0.8132          & 0.9783          & N/A             & 0.9796          \\
            The Proposed Method                                    & 2020 & \textbf{0.9664} & \textbf{0.8374} & 0.9790          & \textbf{0.8127} & \textbf{0.9835} \\ \bottomrule
        \end{tabular}
    }
\end{table}

\begin{table}[!htbp]
    \caption{Comparison with other state-of-the-art methods on the STARE dataset}\label{tab:STARE}
    \resizebox{\columnwidth}{!}{
        \begin{tabular}{@{}llllllll@{}}
            \toprule
            Methods                                                & Year & ACC             & SE              & SP              & F1              & AUROC           \\ \midrule
            Unsupervised Methods                                   &      &                 &                 &                 &                 &                 \\ \midrule
            Mendonca and Campilho\cite{Mendonca:2006dq}            & 2006 & 0.9440          & 0.6996          & 0.9730          & N/A             & N/A             \\
            Al-Diri \textit{et al.}\cite{AlDiri:2009jh}            & 2009 & N/A             & 0.7521          & 0.9681          & N/A             & N/A             \\
            You \textit{et al.}\cite{You:2011do}                   & 2011 & 0.9497          & 0.7260          & 0.9756          & N/A             & N/A             \\
            Fraz \textit{et al.}\cite{Fraz:2012ef}                 & 2012 & 0.9442          & 0.7311          & 0.9680          & N/A             & N/A             \\
            Fathi \textit{et al.}\cite{Fathi:2013di}               & 2013 & N/A             & 0.8061          & 0.9717          & 0.7509          & N/A             \\
            Roychowdhury \textit{et al.}\cite{Roychowdhury:2015da} & 2015 & 0.9560          & 0.7317          & 0.9842          & N/A             & N/A             \\
            Fan \textit{et al.}\cite{Fan:2019dn}                   & 2019 & 0.9570          & 0.7910          & 0.9700          & N/A             & N/A             \\ \midrule
            Supervised Methods                                     &      &                 &                 &                 &                 &                 \\ \midrule
            Staal \textit{et al.}\cite{Staal:2004dd}               & 2004 & 0.9516          & N/A             & N/A             & N/A             & 0.9614          \\
            Marin \textit{et al.}\cite{Marino:2006jf}              & 2011 & 0.9526          & 0.6944          & 0.9819          & N/A             & 0.9769          \\
            Fraz \textit{et al.}\cite{Fraz:2012ef}                 & 2012 & 0.9534          & 0.7548          & 0.9763          & N/A             & 0.9768          \\
            Vega \textit{et al.}\cite{Vega:2015kh}                 & 2015 & 0.9483          & 0.7019          & 0.9671          & 0.6614          & N/A             \\
            Fan \textit{et al.}\cite{Fan:2016kd}                   & 2016 & 0.9588          & 0.6996          & 0.9787          & N/A             & N/A             \\
            Fan and Mo\cite{Fan:2016ie}                            & 2016 & 0.9654          & 0.7834          & 0.9799          & N/A             & N/A             \\
            Liskowski \textit{et al.}\cite{Liskowski:2016}         & 2016 & 0.9729          & 0.8554          & 0.9862          & N/A             & \textbf{0.9928} \\
            Li \textit{et al.}\cite{Li:2016jl}                     & 2016 & 0.9628          & 0.7726          & 0.9844          & N/A             & 0.9879          \\
            Orlando \textit{et al.}\cite{Orlando:2017iy}           & 2017 & N/A             & 0.7680          & 0.9738          & 0.7644          & N/A             \\
            Mo and Zhang\cite{Mo:2017io}                           & 2017 & 0.9674          & 0.8147          & 0.9844          & N/A             & 0.9885          \\
            Xiao \textit{et al.}\cite{Xiao:2018hs}                 & 2018 & 0.9693          & 0.7469          & N/A             & N/A             & N/A             \\
            Alom \textit{et al.}\cite{Alom:2019id}                 & 2019 & 0.9712          & 0.8292          & \textbf{0.9862} & 0.8475          & 0.9914          \\
            Lei \textit{et al.}\cite{Lei:2020}                     & 2019 & 0.9698          & 0.8398          & 0.9761          & N/A             & 0.9858          \\
            The Proposed Method                                    & 2020 & \textbf{0.9713} & \textbf{0.8664} & 0.9798          & \textbf{0.8191} & 0.9875          \\ \bottomrule
        \end{tabular}
    }
\end{table}

\begin{table}[!htbp]
    \caption{Comparison with other state-of-the-art methods on the CHASE\_DB1 dataset}\label{tab:CHASE}
    \resizebox{\columnwidth}{!}{
        \begin{tabular}{@{}llllllll@{}}
            \toprule
            Methods                                                & Year & ACC             & SE              & SP              & F1              & AUROC           \\ \midrule
            Unsupervised Methods                                   &      &                 &                 &                 &                 &                 \\ \midrule
            Fraz \textit{et al.}\cite{Fraz:2012ef}                 & 2012 & 0.9469          & 0.7224          & 0.9711          & N/A             & 0.9712          \\
            Roychowdhury \textit{et al.}\cite{Roychowdhury:2015da} & 2015 & 0.9467          & 0.7615          & 0.9575          & N/A             & N/A             \\
            Fan \textit{et al.}\cite{Fan:2019dn}                   & 2019 & 0.9510          & 0.6570          & 0.9730          & N/A             & N/A             \\ \midrule
            Supervised Methods                                     &      &                 &                 &                 &                 &                 \\ \midrule
            Fraz \textit{et al.}\cite{Fraz:2014fx}                 & 2014 & N/A             & 0.7259          & 0.9770          & 0.7488          & N/A             \\
            Fan and Mo\cite{Fan:2016ie}                            & 2016 & 0.9573          & 0.7656          & 0.9704          & N/A             & N/A             \\
            Liskowski \textit{et al.}\cite{Liskowski:2016}         & 2016 & 0.9628          & 0.7816          & 0.9836          & N/A             & 0.9823          \\
            Li \textit{et al.}\cite{Li:2016jl}                     & 2016 & 0.9527          & 0.7569          & 0.9816          & N/A             & 0.9738          \\
            Orlando \textit{et al.}\cite{Orlando:2017iy}           & 2017 & N/A             & 0.7277          & 0.9712          & 0.7332          & N/A             \\
            Mo and Zhang\cite{Mo:2017io}                           & 2017 & 0.9581          & 0.7661          & 0.9793          & N/A             & 0.9812          \\
            Alom \textit{et al.}\cite{Alom:2019id}                 & 2019 & 0.9634          & 0.7756          & 0.9820          & 0.7928          & 0.9815          \\
            Lei \textit{et al.}\cite{Lei:2020}                     & 2019 & 0.9648          & 0.8275          & 0.9768          & N/A             & 0.9812          \\
            The Proposed Method                                    & 2020 & \textbf{0.9759} & \textbf{0.8670} & \textbf{0.9840} & \textbf{0.8313} & \textbf{0.9905} \\ \bottomrule
        \end{tabular}
    }
\end{table}

\begin{table}[!htbp]
    \caption{Comparison with other state-of-the-art methods on the HRF dataset}\label{tab:HRF}
    \resizebox{\columnwidth}{!}{
        \begin{tabular}{@{}llllllll@{}}
            \toprule
            Methods                                                & Year & ACC             & SE              & SP              & F1              & AUROC           \\ \midrule
            Unsupervised Methods                                   &      &                 &                 &                 &                 &                 \\ \midrule
            Roychowdhury \textit{et al.}\cite{Roychowdhury:2015da} & 2015 & 0.9467          & 0.7615          & 0.9575          & N/A             & N/A             \\ \midrule
            Supervised Methods                                     &      &                 &                 &                 &                 &                 \\ \midrule
            Kolar \textit{et al.}\cite{Odstrcilik:2013ir}          & 2013 & N/A             & 0.7794          & 0.9584          & 0.7158          & N/A             \\
            Orlando \textit{et al.}\cite{Orlando:2017iy}           & 2017 & N/A             & 0.7874          & 0.9584          & 0.7158          & N/A             \\
            The Proposed Method                                    & 2020 & \textbf{0.9698} & \textbf{0.8076} & \textbf{0.9831} & \textbf{0.7963} & \textbf{0.9845} \\ \bottomrule
        \end{tabular}
    }
\end{table}

The computation time needed for processing a fundus image in the DRIVE dataset
using the proposed method is also compared with other state-of-the-art methods
that use patch based or end-to-end approaches and
the results are shown in Table~\ref{tab:time}.
Without the need of cropping and merging patches
as in the patch based UNet\cite{Antiga:2016wf},
the end-to-end approaches are typically faster.
Without the need of pre- and post- processing,
the proposed method can generate high resolution vessel segmentation
in a single forward feeding of a full-sized fundus image
in the DRIVE dataset in about \(0.4s\).

\begin{table}[!htbp]
    \caption{Computation time comparison with other state-of-the-art methods on a fundus image in the DRIVE dataset.}\label{tab:time}
    \resizebox{\columnwidth}{!}{
        \begin{tabular}{@{}llll@{}}
            \toprule
            Methods                                                             & Year & Main device              & Computation time \\ \midrule
            Antiga\cite{Antiga:2016wf}  (patch based UNet)                      & 2016 & NVIDIA GTX Titan Xp      & 10.5 s           \\
            Mo and Zhang\cite{Mo:2017io} (end-to-end FCN)                       & 2017 & NVIDIA GTX Titan Xp      & \textbf{0.4 s}   \\
            Fan \textit{et al.}\cite{Fan:2019dn} (unsupervised image matting)   & 2019 & NVIDIA GTX Titan Xp      & 6.2 s            \\
            Alom \textit{et al.}\cite{Alom:2019id} (patch based recurrent UNet) & 2019 & NVIDIA GTX Titan Xp      & 2.8 s            \\
            The baseline UNet                                                   & 2020 & Intel Xeon CPU @ 2.40GHz & 4.5 s            \\
            The Proposed Method                                                 & 2020 & Intel Xeon CPU @ 2.40GHz & 3.0 s            \\
                                                                                &      & NVIDIA GTX Titan Xp      & \textbf{0.4 s}   \\ \bottomrule
        \end{tabular}
    }
\end{table}

\subsection{Sensitivity analysis of global threshold}\label{subsec:SenAnalysis}
\begin{figure}[!htbp]
    \centering
    \includegraphics[width=\columnwidth]{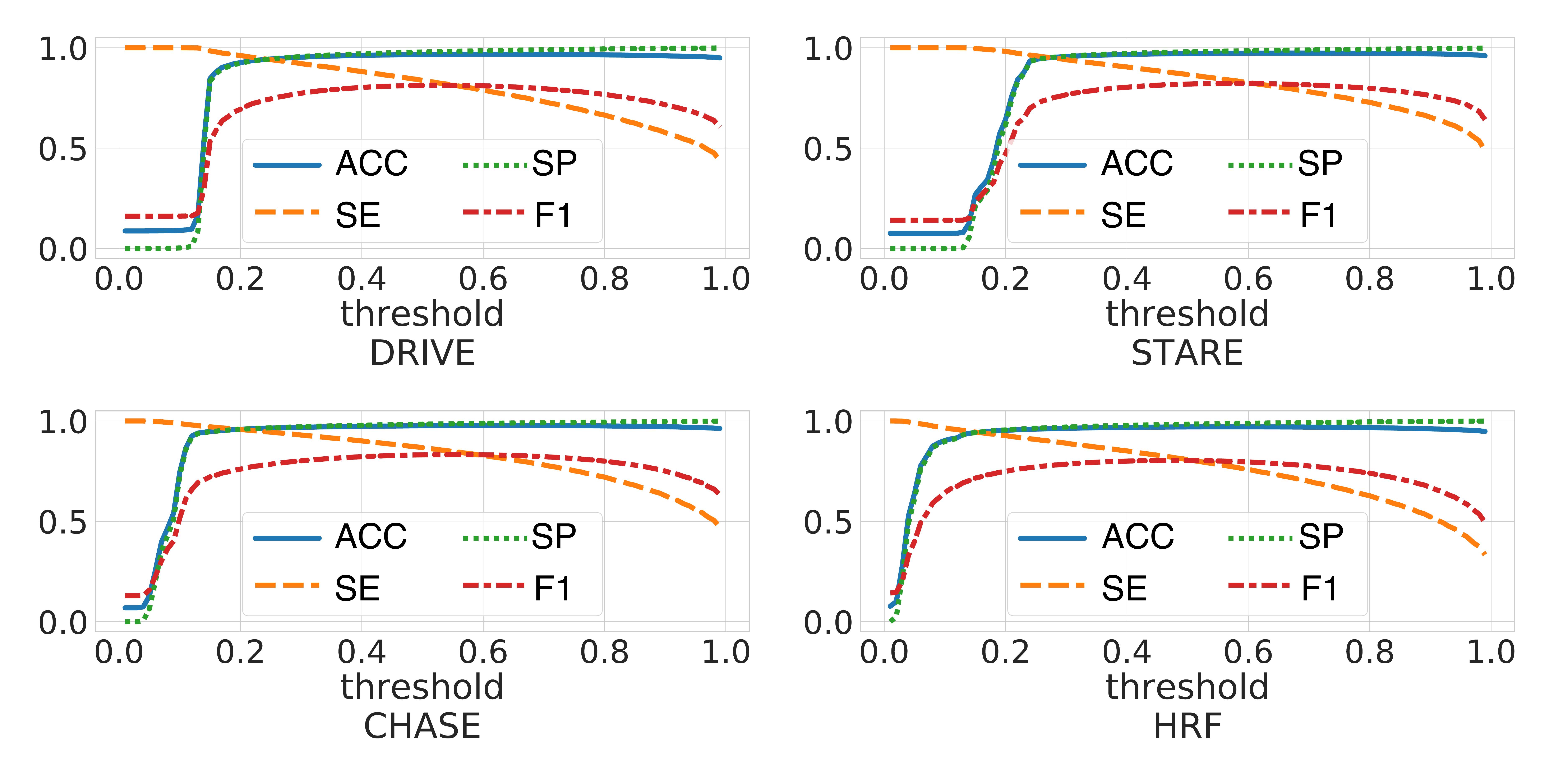}
    \caption{Sensitivity curves of the proposed method on different datasets.}\label{fig:Sen}
\end{figure}

The threshold-sensitive metrics: ACC, SE, SP, and F1
are measured at various global threshold values
sampled in \(\tau_{i} \in 0.01, \dots, 0.99\).
The resulting sensitivity curves are shown in Fig.\ref{fig:Sen}.
The sensitivity curves of the proposed method have
similar profiles on all datasets tested,
which demonstrates the robustness of the proposed Octave UNet
across different datasets.
Furthermore, near the adopted threshold, i.e., \(\tau = 0.5\),
the sensitivity curves of ACC, SP, and F1 change very slightly,
which further demonstrates the robustness
of the proposed method against the global threshold.
Moreover, by lowering the threshold \(\tau \) to about 0.25,
the proposed method can achieve significant gain on SE,
while the other metrics decrease very slightly.

\subsection{Performance on abnormal cases}\label{subsec:cases}
\begin{figure}[!htbp]
    \centering
    \includegraphics[width=.98\columnwidth]{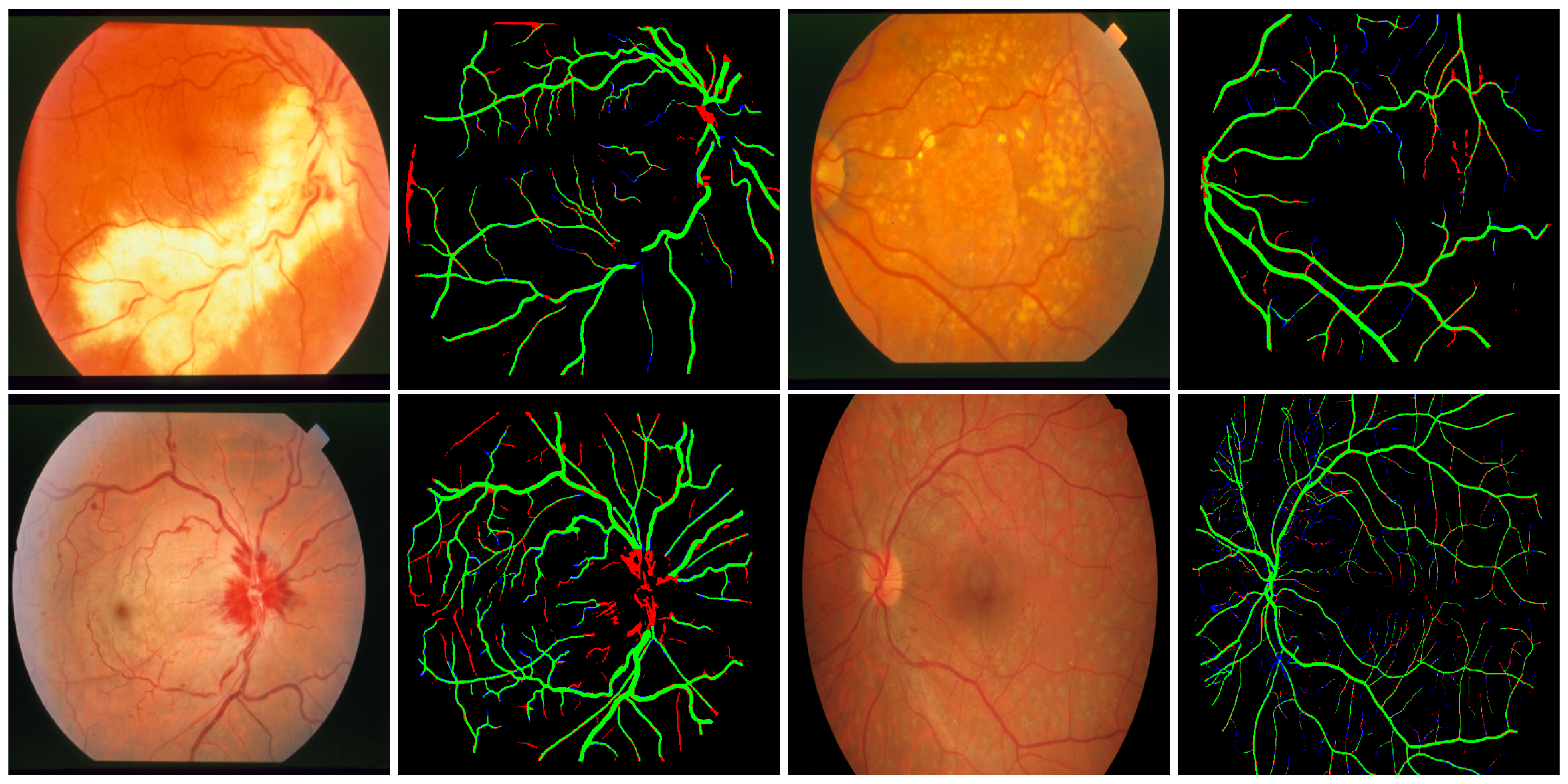}
    \caption{The original fundus images and the analytical vessel segmentation results of cases with exudates, cotton wool spots, hemorrhages, and lesions, respectively.
        The proposed method can segment retinal vasculatures with different types of abnormal regions in fundus images.
        The analytic results (as shown in the second and last columns) are plotted by painting TP with green pixels, FP with red pixels, FN with blue pixels, and TN with black pixels.
    }\label{fig:abnormal}
\end{figure}

The vessel segmentation results of the proposed method
on cases with abnormalities are shown in Fig.\ref{fig:abnormal},
which demonstrates the robust performance of the proposed method
against various abnormalities such as exudates,
cotton wool spots, hemorrhages, and pathological lesions.

\begin{figure}[!htbp]
    \centering
    \includegraphics[width=\columnwidth]{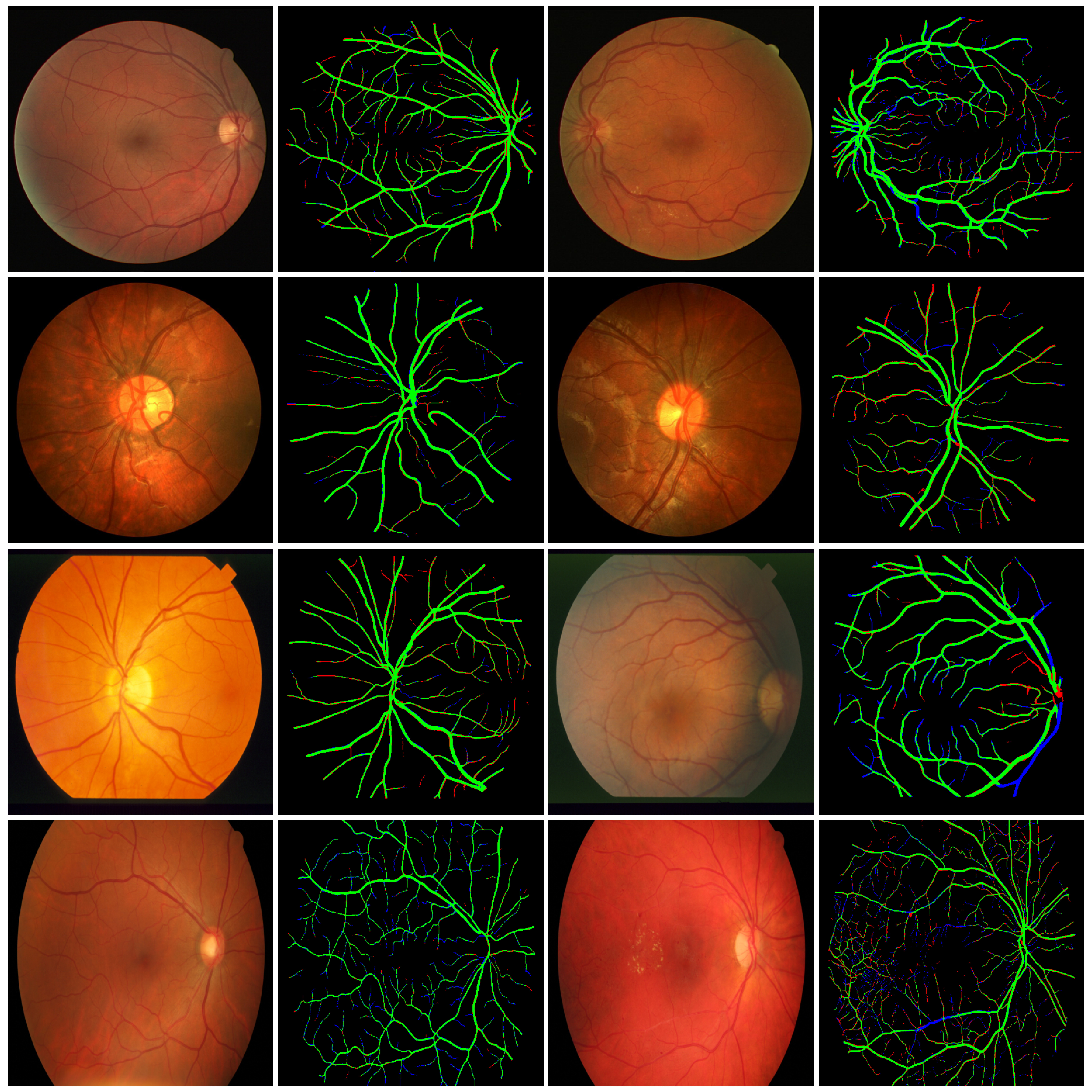}
    \caption{The best (first 2 columns) and worst (last 2 columns) cases on DRIVE (first row),
        STARE (second row), CHASE\_DB1 (third row) and HRF (last row) datasets.
        The analytic results (second and last columns) are obtained by plotting TP with green, FP with read, FN with blue, and TN with black.
    }\label{fig:best-and-worst-cases}
\end{figure}

The best and worst performances on different datasets are illustrated
in Fig.\ref{fig:best-and-worst-cases}.
The best cases on all datasets contain very few missed thin vessels.
On the other hand, the worst cases show that
the proposed method are affected by uneven illuminations.
In both best and worst cases, the proposed method
is capable of discriminatively separating
non-vascular structures and vasculatures.

\section{Conclusion}\label{sec:Conclusion}
An effective and efficient method for retinal vessel segmentation
based on multifrequency convolutional network is proposed in this paper.
Built upon octave convolution and the proposed
octave transposed convolution, the carefully designed Octave UNet can
extract hierarchical features with multiple-spatial-frequencies
and reconstruct accurate vessel segmentation maps.
Benefiting from the design of hierarchical multifrequency features,
the Octave UNet outperforms the baseline model
in terms of both segmentation performances and computational expenditure.
Comprehensive experimental results also demonstrate that the Octave UNet
achieves better or comparable performances
against the other state-of-the-art methods.

\end{document}